\documentstyle{article}
\newcommand{\be}{\begin{equation}}
\newcommand{\ee}{\end{equation}}

\def\bea{\begin{eqnarray}}
\def\eea{\end{eqnarray}}
\def\bean{\begin{eqnarray*}}
\def\eean{\end{eqnarray*}}
\def\thru#1{\mathrel{\mathop{#1\!\!\!\!/}}}
\def\pa#1{\frac{\partial}{\partial z^{}_#1}}
\def\pab#1{\frac{\partial}{\partial \overline z^{}_#1}}
\def\part#1{\partial^{}_#1}
\def\partb#1{\overline\partial^{}_#1}
\begin{document}

\begin{titlepage}
\begin{flushright}    UFIFT-HEP-02-17 \\ 
%hep-ph/0207253 
\end{flushright}
\vskip 1cm
\centerline{\LARGE{\bf {Supersymmetry and  Euler Multiplets}}}
%\vskip .5cm
%\centerline{\LARGE{\bf { and M-Theory }}}

\vskip 1.5cm
\centerline{\bf Lars Brink }
\vskip .5cm
\centerline{\em Department of Theoretical Physics}
\centerline{\em Chalmers University
of Technology, }
\centerline{\em S-412 96 G\"oteborg, Sweden}
\vskip 1cm
\centerline{\bf Pierre Ramond${}^{\,}$\footnote{Supported in part
by the US Department of Energy under grant DE-FG02-97ER41029} and Xiaozhen Xiong${}^{1}$} 
\vskip .5cm
\centerline{\em  Institute for Fundamental Theory,}
\centerline{\em Department of Physics, University of Florida}
\centerline{\em Gainesville FL 32611, USA}
\vskip 1.5cm

\centerline{\bf {Abstract}}
\vskip .5cm
\noindent Some massless supermultiplets appear as the trivial  solution of Kostant's equation, a Dirac-like equation over special cosets. We study two examples; one over the coset  $SU(3)/SU(2)\times U(1)$ contains  the $N=2$ hypermultiplet in $(3+1)$ dimensions with $U(1)$ as helicity; the other over the coset $F_4/SO(9)$ describes the $N=1$ supermultiplet in eleven dimensions, where $SO(9)$ is the 
light-cone little group.  We present the general solutions to Kostant's equation for both cases; they describe  massless physical states of arbitrary spins which  display the same relations as the  fields in the supermultiplets. They come in sets of three representations called Euler triplets, but do not display supersymmetry although the number of bosons and fermions is the same when spin-statistics is satisfied.   We build the  free  light-cone Lagrangian for both cases.  

\vfill
\begin{flushleft}
July  2002 \\
\end{flushleft}
\end{titlepage}

\section{Introduction}
String theory has given us an understanding of the dimension of spacetime as a crucial concept for building consistent 
quantum mechanical theories. This insight has been  further strengthened with the introduction of supersymmetry, since
the  spinor depends not only on the  spacetime dimensions but also
on the possibility to implement the Majorana and Weyl conditions.
The little group for the superstring is $SO(8)$, one of the
most beautiful and unique groups. Its Dynkin diagram is the famous Mercedes symbol,
and the group is really the Mercedes of the orthogonal groups. It has a triality symmetry and its three eight-dimensional  vector, 
spinor and  cospinor representations are readily interchangeable. In the light-cone
formulation of the superstring it is these three representations that build up the theory and in quantum calculations there are marvellous
cancellations between the bosonic and fermionic contributions that render the theory perturbatively finite.

From this viewpoint the emergence of M-theory as an even more general
theory was unexpected.  Eleven-dimensional space is singled out as the maximal one to carry a supermultiplet
with the graviton as the highest spin field, but is there some group-theoretic reasons why an
eleven-dimensional spacetime should make special sense? The little group  $SO(9)$ has a nondescript  Dynkin diagram. It displays no symmetry and  no cars use it as a symbol, not even the Trabant. 

In this paper, we show that $SO(9)$ does have hidden attributes. They come in the form of a wealth of internal relations 
between sets of its representations, again making it a plausible candidate for a fundamental symmetry. It was found~\cite{PR} 
that some of its irreps naturally group together into triplets which are such that bosonic and fermionic degrees of freedom match up the same way as they do in eleven-dimensional supergravity. This makes the triplets interesting not only from a
mathematical point of view but also from a quantum physical one. Higher order loop calculations in a supersymmetric theory have huge
cancellations between  fermionic and  bosonic contributions since their contributions can be written in terms of group-theoretic indices
related to the little group which match up. In eleven dimensional supergravity, the matching among its three fields is not total, but fails in the highest eighth order invariant, which produces a nonrenormalizable divergence in high loop order~\cite{CURTRIGHT}. Each Euler triplet displays the same partial matching among its representations. It is our hope that total matching can be restored only after we sum over an  infinite number of Euler triplets. This would entail the vanishing of an infinite sum of  positive numbers. 

All other Euler triplets involve higher spin massless fields. This is reminiscent of a problem in string
theory which has not been solved, namely the limit of infinite Regge slope, i.e. the zero-tension limit, where one expects all states to become massless, with an infinite number of states for each spin. However this limit is not tractable at the amplitude level. Can we connect it to a theory of massless spins in higher dimensions?

An interesting question in this context is what happens to the symmetries of a massive theory when all particles become massless.
Usually the linearly realized symmetry is enhanced as it happens when a spontaneously broken symmetry is restored. The gauge invariance is of course restored but what happens to the supersymmetry? In a previous paper~\cite{US}, we have shown that a continuous spin representation of the SuperPoincar\'e group connects massless states of arbitrarily high spins. Another alternative, the use of a tower of Euler triplets, seems to require much bigger  supersymmetries. This issue will be discussed in this paper.  

Since the discovery of the triplets there has been great progress in the
mathematical understanding of their beautiful properties \cite{GKRS}. They arise 
for embeddings where both group and subgroup have the same rank. In the case above $SO(9)$, alias $B_4$, is a subgroup
of $F_4$ with the same rank, and the quotient space $F_4/B_4$ has Euler
number three giving a triplet of $SO(9)$ to every irrep of  $F_4$.   We have elsewhere \cite{BR,BR2,PR2}
listed the cases with up to 16-dimensional cosets and in this list we
find multiplets with the above properties, extending the multiplets of
besides the 11-dimensional supergravity also  $N=8$ supergravity, $N=4$ Yang-Mills
and the $N=2$ hypermultiplet. All these Euler triplets arise as solutions of Kostant's equation~\cite{KOS}, which is a Dirac-like equation on the coset.

It is quite interesting that the exceptional algebra $F_4$ enters into
the description of an 11-dimensional theory. We have seen the
exceptional groups emerge as gauge groups and there is a direct line
from the gauge group of the Standard Model via $SU(5) = E_4$ and $SO(10)
= E_5$ up to the ultimate exceptional group $E_8$, the Delahaye of the groups. (Rolls Royce for non-French readers.) So far there has not
been any trace of exceptional groups extending the space time symmetry.
There is a simple reason for this since they relate tensor and spinor
representations of their orthogonal subgroups, while spin statistics
treat them differently. However, the exceptional groups are the most
unique and beautiful ones and it is many physicists' dream that they represent
the ultimate symmetry of the world. 

One question that will arise is if we should use all triplets as a candidate
theory or if there is a natural selection among them. In this report we show a method where we define a superfield which naturally has an
equal number of bosonic and fermionic degrees of freedom and is a
natural extension of the superfield containing the smallest multiplet. We will use a light-cone  frame formulation which
will tie together the superspace with the external symmetry. 

We start by  discussing  in  some detail the simpler case of $SU(3)/SU(2) \times U(1)$, which we interpret as a model for higher spin massless fields in four dimensions, generalizing the $N=2$ hypermultiplet. We then apply the same methods to the more complicated case of more direct physical interest where the Euler triplets appear as a natural generalization of supergravity in eleven dimensions.

\setcounter{equation}{0}
\section{Euler Triplets for $SU(3)/SU(2)\times U(1)$}
We now present a detailed analysis of the Euler triplets 
associated with the coset $SU(3)/SU(2)\times U(1)$. There are infinitely many of them, one for each representation of $SU(3)$. 
The  trivial solution describes the light-cone degrees of freedom of the $N=2$ hypermultiplet in four dimensions, with $U(1)$ interpreted as helicity. Hence we begin by reminding the reader of the well-known light-cone description of that multiplet.  

\subsection{The $N=2$ Hypermultiplet in 4 Dimensions}
The massless $N=2$ scalar hypermultiplet contains two Weyl spinors and two complex scalar fields, on which   the $N=2$ 
SuperPoincar\'e algebra is realized. Introduce the  light-cone Hamiltonian

\be
P_{}^-=\frac{p\overline p}{p^+}\ ,\ee
where 
$
p=\frac{1}{\sqrt{2}}(p_{}^1+ip_{}^2)\nonumber \ .$ The front-form 
supersymmetry generators satisfy the anticommutation relations

\bea \nonumber \{{\cal Q}^{m}_+ ,\overline {\cal Q}^{n}_+\}&=&-
2\delta_{}^{mn}p_{}^+\ ,\\
\{{\cal Q}^{m}_- ,\overline {\cal Q}^{n}_-\}&=&-2\delta_{}^{mn}
\frac{p\overline p}{p_{}^+}\ ,~~~~~m,n=1,2\ ,\\
\nonumber \{{\cal Q}^{m}_+ ,\overline {\cal Q}^{n}_-\}&=&-2\delta_{}^{mn}p\ .\eea
The kinematic supersymmetries are expressed as 

\be
{\cal Q}^{m}_+=-\frac{\partial}{\partial\overline\theta^m}-\theta_m p_{}^+\ ,\qquad
\overline {\cal Q}^{m}_+=\frac{\partial}{\partial\theta^m}+\overline\theta_m p_{}^+\ ,\ee
while the kinematic Lorentz generators are given by

\bea
\label{M}
M_{}^{12}&=&i(x\overline p-\overline xp)+\frac{1}{2}\theta_m
\frac{\partial}{\partial\theta_m}-\frac{1}{2}\overline\theta^m
\frac{\partial}{\partial\overline\theta^m}\ ,\\ \nonumber   
M_{}^{+-}&=&-x_{}^-p_{}^+ -\frac{i}{2}\theta_m\frac{\partial}
{\partial\theta_m}-\frac{i}{2}\overline\theta^m\frac{\partial}
{\partial\overline\theta^m}\ ,   
\\ \nonumber
M_{}^{+}&\equiv&\frac{1}{\sqrt{2}}(M_{}^{+1}+iM_{}^{+2})=
-xp_{}^+\ ,\qquad 
\overline M_{}^{+}=-\overline xp_{}^+\ ,\\ \nonumber
\eea
where $x=\frac{1}{\sqrt{2}}(x_{}^1+ix_{}^2)$, and where the two complex Grassmann variables satisfy
the anticommutation relations
\bea
\nonumber \{\theta_m,\frac{\partial}{\partial\theta_n} \}&=&
\{\overline\theta^m,\frac{\partial}{\partial\overline\theta^n}\}=\delta^{mn}\ ,\\ 
\nonumber \{\theta_m,\frac{\partial}{\partial\overline\theta^n} \}&=&
\{\overline\theta^m,\frac{\partial}{\partial\theta_n} \}
=0\ .\eea
The  (free) Hamiltonian-like supersymmetry generators are simply
\be
{\cal Q}^{m}_-=\frac{\overline p}{p_{}^+}{\cal Q}^{m}_+\ ,\qquad
\overline {\cal Q}^{m}_-=\frac{ p}{p_{}^+}\overline {\cal Q}^{m}_+\ ,\ee
and the light-cone boosts are given by
\bea
M_{}^{-}&=&x_{}^-p-\frac{1}{2}\{x,P_{}^-\}+i\frac{p}{p_{}^+}\theta_m
\frac{\partial}{\partial\theta_m}\ ,
\\ \nonumber
\overline M_{}^{-}&=&x_{}^-\overline p-\frac{1}{2}\{\overline x,P_{}^-\}+i\frac{
\overline p}{p_{}^+}\overline \theta^m
\frac{\partial}{\partial\overline \theta^m}\ .\eea
This representation of the superPoincar\'e algebra is reducible, as it 
can be seen to act on reducible superfields $\Phi(x^-,x^i,\theta_m,\overline\theta^m)$, because  the operators 

\be
{\cal D}_+^m~=\frac{\partial}{\partial\overline\theta^m}-\theta_m p_{}^+\ ,\ee
 anticommute with the supersymmetry generators. As a result, one can 
achieve irreducibility by acting on superfields for which 
\be
\label{Ch}
{\cal D}_+^m~\Psi=[\frac{\partial}{\partial\overline\theta^m}-\theta_m p_{}^+]\Psi =0\ ,\ee
solved by the chiral superfield

\be
\label{Phi}
\Psi(y^-,x^i,\theta_m)= \psi^{}_0(y^-,x^i) + \theta^{}_m \psi_{}^m(y^-,x^i) + 
\theta^{}_1 \theta^{}_2\psi_{}^{12}(y^-,x^i)\ .\ee
The field entries of the scalar hypermultiplet now depend on the combination

\be
y^-=x^--i\theta_m\overline\theta^m\ ,\ee
and the transverse variables.  Acting on this chiral superfield, the 
constraint is equivalent to requiring that

\be
{\cal Q}_+^m\approx -2p^+\theta_m\ ,\qquad \overline{\cal
Q}_+^m\approx 
\frac{\partial}{\partial\theta_m}\ ,\ee
where the derivative is meant to act only on the naked $\theta_m$'s, 
not on those hiding in  $y^-$. This light-cone representation is 
well-known, but we repeat it here to set our conventions and notations.

\subsection{Kostant's Equation }
Let $T^A$ , $A=1,2,\dots 8$, denote the  $SU(3)$ generators. Its $SU(2)\times U(1)$
subalgebra is generated by  $T^i$, $i=1,2,3$,  and $T^8$. Introduce  Dirac matrices over the coset 

$$ \{\gamma^a, \gamma^b \} = 2\delta ^{ab}\ ,$$
for $a,b=4,5,6,7$, to define 
 the Kostant equation over the coset $SU(3)/SU(2)\times U(1)$ as

\be
\thru {\cal K}\Psi~=~\sum_{a=4,5,6,7}\gamma_{}^aT^{}_a\,\Psi~=~0\ .\nonumber \ee
 The Kostant operator commutes with  the $SU(2)\times U(1)$ generators

\be
L^{}_i=T^{}_i+S^{}_i\ ,~i=1,2,3\ ;\qquad L^{}_8=T^{}_8+S^{}_8\ ,\ee
 sums of the $SU(3)$ generators and of the $``$spin" part, expressed in terms 
of the $\gamma$ matrices as 

\be 
S^{}_j=- \frac{i}{4}f^{}_{jab}\gamma_{}^{ab}\ , \qquad S^{}_8=-
\frac{i}{4}f^{}_{8ab}\gamma_{}^{ab}\ ,\ee
where $\gamma^{ab}=\gamma^a\gamma^b\ ,a\ne b\ ,$ and $f^{}_{jab}\ ,f^{}_{8ab}$ are structure functions of $SU(3)$.

The Kostant equation has an infinite number of solutions which come in groups of three representations of 
$SU(2)\times U(1)$, called Euler triplets. For each representation of
$SU(3)$, there is a unique Euler triplet, each given by three representations

$$\{a_1,a_2\}~\equiv~[a^{}_2]_{-\frac{2a^{}_1+a_2+3}{6}}\oplus 
[a^{}_1+a_2+1]_{\frac{a^{}_1-a_2}{6}}\oplus [a^{}_1]_{\frac{2a^{}_2+a_1+3}{6}}\ ,$$
where $a_1,a_2$ are the Dynkin labels of the associated $SU(3)$ representation. 
Here, $[a]$ stands for the $a=2j$ representation of $SU(2)$, and the
subscript denotes the $U(1)$ charge. The Euler triplet corresponding to $a_1=a_2=0$,

$$\{0,0\}~=~[0]_{-\frac{1}{2}}\oplus [1]_{0}\oplus [0]_{\frac{1}{2}}\ ,$$
describes the degrees of freedom of the $N=2$ supermultiplet,
where the properly normalized $U(1)$ is interpreted as the helicity of the four-dimensional Poincar\' e algebra. 

Below, we wish to explore the  possibility of linking this 
supersymmetric triplet to those for which $a_{1,2}\ne 0$, 
while preserving at least relativistic invariance. Of particular interest will be the algebraic operations that link the different Euler triplets. Their use will enable us to define supersymmetry-like  operations acting on the higher Euler triplets, which serve as the shadow of the light-cone supersymmetry of the lowest Euler triplet.

The $U(1)$ 
charges of the higher triplets are  generally rational numbers, which means that they display parastatistics, but the triplets for which  

$$a_1~=~a_2~~{\rm mod}~(3)\ ,$$
contain half-odd integer or integer $U(1)$ charges, and satisfy Fermi-Dirac statistics. A self-conjugate subset  

$$\{a,a\}~:~~~~[a]_{-\frac{a+1}{2}}\oplus [2a+1]_{0}\oplus [a]_{\frac{a+1}{2}}\ ,$$
 contains equal number of half-odd integer-helicity fermions and
integer-helicity bosons, and satisfy CPT.  As the helicity gap between the representations increases indefinitely in half integer steps, the symmetry operations that relate its members have helicities $\pm(a+1)/2$. When $a=0$, they can be identified with the usual supersymmetries, and they are fermionic as long as $a$ is even. The others are generated from  complex representations with  $N=1,2,\dots$ like

$$\{a,a+3N\}~:~~~~[a]_{-\frac{a+N+1}{2}}\oplus
[2a+3N+1]_{\frac{N}{2}}\oplus [a+3N]_{\frac{a+N+1}{2}}\ .$$ 
The helicity gap also increases in half integer steps, starting at one-half. 
Since  CPT requires states of opposite helicity, these must be 
accompanied by their conjugates, 
 $\{a+3N,a\}$, with all helicities reversed. 

A special case deserves consideration: when $a=0$, the helicity gap can be as small as $1/2$, like the regular supersymmetry. The simplest example is

$$\{0,3\}~:~~~~[0]_{-1}\oplus
[4]_{\frac{1}{2}}
\oplus [3]_{1}\ ,$$
where the helicity gap is $3/2$ and $1/2$. When we add the $CPT$ conjugate

$$ \{3,0\}~:~~~~   \oplus [0]_{1}\oplus
[4]_{-\frac{1}{2}}
\oplus [3]_{-1}\ ,$$
  we end up with states separated by half a unit of helicity as in the supersymmetric multiplets. As they occur in different representations of $SU(2)$,  equality between bosons and fermions is achieved only after including the CPT conjugate, but as long as $SU(2)$ remains unbroken, relativistic supersymmetry cannot be implemented on these states. The case $N=2$ yields states of helicity $1$ and $3/2$, and $N=3$ contains eleven states of helicity $2$, and so on.

It appears that while Poincar\'e symmetry can be implemented on an infinite subset of Euler triplets, relativistic supersymmetry can be realized on a finite subset, and only after the $SU(2)$ is broken. In particular the need for operators that shift helicity by more than half units makes it unlikely that a relativistic supersymmetric theory of Euler triplets can be found. 

In addition, the higher Euler triplets include states with  helicities larger than $2$, which cannot be interpreted 
as massive relativistic states since they do not arrange themselves in $SO(3)$ representations. A relativistic description requires them to be  massless particles of spin higher than $2$ in four dimensions.

There are well-known difficulties with such theories~\cite{DIFFICULT}; in the flat space limit they must decouple from the gravitational 
sector, although this can be circumvented in curved space~\cite{VASILIEV}, or if   there 
is an infinite number of such particles.  
Our purpose is to investigate if a relativistic theory can be formulated 
with an infinite number of Euler multiplets, in which a light-cone version of a new type of space-time fermionic symmetry is present. 

\subsection{Grassmann Numbers and Dirac Matrices}
We will express the solutions of Kostant's equations as chiral superfields over space-time and internal variables pertaining to $SU(3)$. For that purpose,  we first identify the spin part of the $U(1)$ generator $S^{}_8$ with the spin part in Eq.~(\ref{M}) taking the condition (\ref{Ch}) into account, thus  writing the $S^{}_i$ in terms of  $\theta's$. An appropriate representation is then

\bea
\gamma_{}^4+i\gamma_{}^5&=&i\sqrt{\frac{2}{p^+}}{\cal Q}_+^1\ ,\qquad
\gamma_{}^4-i\gamma_{}^5=i\sqrt{\frac{2}{p^+}}\overline{\cal Q}_+^1\\
\gamma_{}^6+i\gamma_{}^7&=&i\sqrt{\frac{2}{p^+}}{\cal Q}_+^2\ ,\qquad
\gamma_{}^6-i\gamma_{}^7=i\sqrt{\frac{2}{p^+}}\overline{\cal Q}_+^2\ ,
\eea
in terms of the kinematic $N=2$ light-cone supersymmetry generators 
defined in the previous section. We can check that $S_8$ indeed agrees with the spin part of Eq.~(\ref{M}) (after proper normalization). As the Kostant operator 
anticommutes with the constraint operators

\be\{~\thru {\cal K},~{\cal D}_+^m\}~=~0\ ,\ee
 its solutions can be written as chiral superfields, on which the $\gamma$'s become   

\bea
\gamma_{}^4+i\gamma_{}^5&=&-2i\sqrt{2p^+}~\theta_1\ ,\qquad
\gamma_{}^4-i\gamma_{}^5=i\sqrt{\frac{2}{p^+}}~\frac{\partial}{\partial\theta_1}\\
\gamma_{}^6+i\gamma_{}^7&=&-2i\sqrt{2p^+}~\theta_2\ ,\qquad
\gamma_{}^6-i\gamma_{}^7=i\sqrt{\frac{2}{p^+}}~\frac{\partial}{\partial\theta_2}\ ,
\eea
The complete $``$spin" parts of the $SU(2) \times U(1)$ generators,  expressed
in terms of  Grassmann variables, do not depend on $p^+$,

\bea
S^{}_1&=&\frac{1}{2}  ( \theta^{}_1\frac{\partial}{\partial
\theta^{}_2}+ 
\theta^{}_2\frac{\partial}{\partial \theta^{}_1})\ ,\qquad
S^{}_2=-\frac{i}{2}  (\theta^{}_1 \frac{\partial}{\partial
\theta^{}_2}-  
\theta^{}_2 \frac{\partial}{\partial \theta^{}_1})\cr
S^{}_3&=&\frac{1}{2}  (\theta^{}_1\frac{\partial}{\partial
\theta^{}_1}-  
\theta^{}_2\frac{\partial}{\partial \theta^{}_2})\ ,\qquad
S^{}_8=\frac{\sqrt3}{2} (\theta^{}_1 \frac{\partial}{\partial
\theta^{}_1}+
\theta^{}_2 \frac{\partial}{\partial \theta^{}_2}-1)\ .
\eea
Using  Grassmann properties, the $SU(2)$ Casimir operator can be written as  

\be\vec{S}^2= \frac{3}{4}  ( \theta^{}_1\frac{\partial}{\partial
\theta^{}_1}  
- \theta^{}_2 \frac{\partial}{\partial \theta^{}_2})^2\ ;\ee
it has only two eigenvalues, $3/4$ and zero. These $SU(2)$ generators obey a simple algebra

\be
S^{}_i\,S^{}_j=\frac{1}{3}\vec S\cdot \vec S\,\delta^{}_{ij}+\frac{i}{2}\epsilon^{}_{ijk}S^{}_k\ .\ee
The helicity, identified with $S_8$ up to a normalizing factor
of $\sqrt{3}$, leads to half-integer helicity values on the
Grassmann-odd components of the (constant) superfield representing the hypermultiplet. 

\subsection{Solutions of Kostant's Equation}
Consider now  Kostant's equation over $SU(3)/SU(2)\times U(1)$. It is given by

$$\thru {\cal K}\Psi~=~\sum_{a=4,5,6,7}\gamma_{}^aT^{}_a\,\Psi~=~0\ ,$$
where $\Psi$ is now a chiral superfield as in (9) extended to contain internal group variables. Expanding the solutions and the Dirac matrices in terms of Grassmann 
variables yields two independent pairs of equations

$$(T^{}_4+iT^{}_5)\psi^{}_1+(T^{}_6+iT^{}_7)\psi^{}_2~=~0\ ;\qquad 
(T^{}_4-iT^{}_5)\psi^{}_2-(T^{}_6-iT^{}_7)\psi^{}_1~=~0\ ,$$
 and

$$(T^{}_4-iT^{}_5)\psi^{}_0-(T^{}_6+iT^{}_7)\psi^{}_{12}~=~0\ ;\qquad 
(T^{}_6-iT^{}_7)\psi^{}_0+(T^{}_4+iT^{}_5)\psi^{}_{12}~=~0\ ,$$
that is in the representation of the generators given in Appendix A

$$(z^{}_1\part3-\overline z^{}_3\partb1)\psi^{}_1+(z^{}_2\part3-
\overline z^{}_3\partb2)\psi^{}_2~=~0\ ;\qquad (z^{}_3\part1-
\overline z^{}_1\partb3)\psi^{}_2-(z^{}_3\part2-\overline z^{}_2
\partb3)\psi^{}_1~=~0\ ,$$
$$(z^{}_3\part1-\overline z^{}_1\partb3)\psi^{}_0-(z^{}_2\part3-
\overline z^{}_3\partb2)\psi^{}_{12}~=~0\ ;\qquad (z^{}_3\part2-
\overline z^{}_2\partb3)\psi^{}_0+(z^{}_1\part3-\overline z^{}_3\partb1)\psi^{}_{12}~=~0\ .$$
The homogeneity operators

$$D=z^{}_1\part1+z^{}_2\part2+z^{}_3\part3\ ,\qquad \overline 
D=\overline z^{}_1\partb1+\overline z^{}_2\partb2+ \overline z^{}_3\partb3\,$$
commute with $\thru {\cal K}$, allowing the solutions of  Kostant
equation to  be arranged 
in terms of homogeneous polynomials, on which  $a_1$ is the eigenvalue of $D$ and $a_2$ that of $\overline D$.
The solutions can also be labeled in terms of the $SU(2)\times U(1)$ generated by the operators

$$L^{}_i=T^{}_i+S^{}_i\ ,~i=1,2,3\ ;\qquad L^{}_8=T^{}_8+S^{}_8\ .$$
The solutions for each triplet, are easily written for the highest weight states of each representation,  

\bea 
\Psi&=& z_3^{a_1}~\overline z_2^{a_2}~~~~~~~~~~{\rm labels}~~[a_2]_{-\frac{2a^{}_1+a_2+3}{6}}
\ ,\cr 
&+& \theta_1~z_1^{a_1}~\overline z_2^{a_2}~~~~~~{\rm labels}~~
[a_1+a_2+1]_{\frac{a^{}_1-a_2}{6}}\ ,\cr
&+& \theta_1\theta_2~z_1^{a_1}~\overline z_3^{a_2}\ ,~~~~{\rm
labels}~
~[a_1]_{\frac{2a^{}_2+a_1+3}{6}}\ ,
\eea
where $[\dots]$ are the  $SU(2)$ Dynkin labels. All other states are obtained by repeated 
action of the lowering operator 

$$L_1-iL_2= \theta^{}_2\frac{\partial}{\partial \theta^{}_1}+ (z^{}_2\part1-\overline z^{}_1\partb2)\ ,$$
giving us all the states within each  Euler triplet. Every state has finally to be multiplied by a field depending on the coordinates $y^{-}$ and $x^i$. We see that in this notation, each triplet differs from the next by the degree of homogeneity. This implies that it is possible to move across the triplets by simple multiplication or differentiation. This is the object of the next section. 

\subsection{Relating the Triplets--Shadow Supersymmetry}
 Since all Euler triplet  superfields can be written in terms of components which are  homogeneous polynomials of order $a_1$ in the variables $\bf z$ and $a_2$ in the $\bf \overline z$, we can easily construct algebraic operations which, when applied to one Euler triplet, generate another. These can then be used to define new operations  called {\it shadow supersymmetry}, which act as ladder operators between the components of a given higher Euler triplet. They generalize the familiar supersymmetry operations to the higher Euler triplets, but  cannot be upgraded to space-time supersymmetry. 

Introduce  the projection operators  onto the components of the chiral superfields:

\bea\nonumber
P^{}_0&=&(1-{\cal P}^{}_1)(1-{\cal P}^{}_2)    \ , 
\qquad
P^{}_1~=~{\cal P}^{}_1(1-{\cal P}^{}_2)\
,\\ \nonumber
P^{}_2&=&{\cal P}^{}_2(1-{\cal P}^{}_1)\
,\qquad
P^{}_{12}~=~{\cal P}^{}_1{\cal P}^{}_2 \ ,\\ \nonumber
\eea
where ${\cal P}_i\equiv \theta_i\partial/\partial\theta_i$, $i=1,2$, and which 
satisfy the relations

\be P_aP_b=\delta_{ab}P_a\ .\ee

There are two basic ladder operators which increase $a_1$ and $a_2$ by one unit, respectively. Let us begin by constructing the spectrum-generating ladder  operators which generate  $\Psi_{\{a_1+1,a_2\}}$ when applied to $\Psi_{\{a_1,a_2\}}$. Here we limit our construction of 
these operations to the highest weight states of each Euler triplet, and refer the reader to Appendix B for a more general presentation.  

Since the non-Grassmann component for each triplet differs simply by the power in $z_3$, we start with  the  operator

\be {\cal A}^{\dagger}_{}~\equiv~z^{}_3~P^{}_0\ ,\ee
which multiplies the lowest component of any superfield by
$z_3$. An $SU(2)$ singlet with  $-1/3$ helicity, it  acts
as a raising operator between Euler triplets. 
The corresponding lowering operator is

\be {\cal A}~\equiv~~P^{}_0\frac{\partial}{\partial z^{}_3}\ .\ee
These two operators act as harmonic oscillators on the lowest
superfield component

\be
[\,{\cal A}~,~{\cal A}^\dagger_{}\,]~=~P^{}_0\ .\ee
The highest weight $\theta$-dependent terms of  two adjacent Euler triplets 
differ  by a single power of $z_1$. This leads us to the operators

$${\cal B}^\dagger_{}~\equiv~z_1\,(1-P^{}_0)\ ,\qquad
{\cal B}^{}_{}~\equiv~\frac{\partial}{\partial z_1}\,(1-P^{}_0)\ ,
 $$
which act as harmonic oscillators on the $\theta$-dependent terms

\be
[\,{\cal B}~,~{\cal B}^\dagger_{}\,]~=~(1-P^{}_0)\ .\ee
It follows that the action of ${\cal A}_{}^{\dagger}$ and ${\cal B}^\dagger_{}$ on the highest weight components of $\Psi_{\{a_1,a_2\}}$ generates the highest weights of the $\Psi_{\{a_1+1,a_2\}}$ Euler triplet. All other states are obtained by the action of the lowering operators. 

A similar construction  for the  ladder operators which change $a_2$ by one unit leads us to the operators

\be {\overline {\cal A}}^{\dagger}_{}~\equiv~P^{}_{12}\,{\bar z}^{}_3 \ ,\qquad  
{\overline {\cal A}}~\equiv~P^{}_{12}~{\frac{\partial}{\partial{\bar z}^{}_3}} \, 
\ee
and 

\be 
{\overline {\cal B}}^\dagger_{}~\equiv~ (1-P^{}_{12})\, {\overline z}^{}_2\ ,\qquad
{\overline {\cal B}}~\equiv~ (1-P^{}_{12})~{\frac{\partial}{\partial {\bar z}^{}_2}} \ ,\ee
which act as harmonic oscillators on the appropriate superfield components.

The action of kinematic supersymmetry among the highest weight components is simply multiplication and differentiation by $\theta_i$. However with the ladder operations we have just introduced, there is another way to move within each triplet. 

For example on the $\Psi_{\{1,0\}}$ superfield, we can use the ladder operator to step down once to the hypermultiplet, perform a supersymmetry operation and then step up once again. This operation,   called  {\em shadow supersymmetry}, is generated by 

\be
{\cal Q}^{\downarrow}_{1}~\equiv~ {\cal B}^\dagger_{}~\theta^{}_1~
{\cal A}\ ,\qquad
{\cal Q}^{\downarrow}_{2}~\equiv~ {\cal B}^\dagger_{}~\theta^{}_2~
{\cal B}\ .\ee
The inverse operations will be

\be
{\cal Q}^{\uparrow}_{1}~\equiv~ {\cal A}^\dagger_{}~\frac{\partial}{\partial \theta^{}_1}~
{\cal B}\ ,\qquad
{\cal Q}^{\uparrow}_{1}~\equiv~ {\cal B}^\dagger_{}~\frac{\partial}{\partial \theta^{}_2}~
{\cal B}\ .
\ee
Properly normalized, these anticommute to one on the ${\{1,0\}}$ superfield. Generalizing to 
 the $\{a_1,a_2\}$ Euler triplet, we can define similar operators

\be
{\cal Q}^{\downarrow\,[a_1,a_2]}_{1}~\equiv~ \left({\cal B}^\dagger_{}\right)^{a_1}_{}\,
\left({\overline{\cal B}}^\dagger_{}\right)^{a_2}_{}~\theta^{}_1~
\left({\cal A}\right)^{a_1}_{}\,\left({\overline{\cal B}}\right)^{a_2}_{}\ ,\ee

\be
{\cal Q}^{\downarrow\,[a_1,a_2]}_{2}~\equiv~ \left({\cal B}^\dagger_{}\right)^{a_1}_{}\,
\left({\overline{\cal A}}^\dagger_{}\right)^{a_2}_{}~\theta^{}_2~
\left({\cal B}\right)^{a_1}_{}\,\left({\overline{\cal B}}\right)^{a_2}_{}\ ,\ee
together with their respective raising counterparts. 
The operators form an anticommuting algebra as they manifestly close on the highest weight components of the Euler triplets, and generalize kinematic supersymmetry. However unlike the usual supersymmetry which change helicity by half a unit, these new operations carry  arbitrarily high values of helicity and cannot be used to construct space-time supersymmetry.

In this section, we have considered the baby example associated with the $SU(3)/SU(2)\times U(1)$ coset, in order to gain familiarity with its Euler triplets. We have built its explicit Euler triplet solutions, and shown  how they are related to each other. While the triplets show no space-time supersymmetry, we have been able to relate its components by nilpotent operations which serve as a generalization of kinematic supersymmetry to the higher triplets. In the next section, we apply the same techniques to the more complicated but much more interesting case of 
$F_4/SO(9)$.

\section{Supergravity in Eleven Dimensions}
The ultimate field theory without gravity is the finite $N=4$ Super Yang-Mills theory in four dimensions~\cite{LJJ}. Eleven dimensional $N=1$ Supergravity~\cite{ELEVEN}, the ultimate field theory with gravity, is not renormalizable; it does not stand on its own as a 
physical theory,   as far as we know today. 
 However, the  eleven-dimensional theory has been recently revived as the infrared limit of the presumably finite M-theory which, like characters on the walls of Plato's  cave, has revealed itself only through its compactified versions onto lower-dimensional manifolds. 

\subsection{SuperAlgebra}
$N=1$ supergravity in eleven dimension is a local field theory that contains three  massless fields, the familiar symmetric second-rank tensor, $h^{}_{ \mu\nu}$ which represents gravity,  a three-form field $A^{}_{\mu\nu\rho}$, and the Rarita-Schwinger spinor $\Psi^{}_{\mu\,\alpha}$.  From its Lagrangian, one can derive the expression for the super Poincar\'e algebra, which in the light cone gauge assumes the particularly simple form in terms of the nine $(16\times 16)$ $\gamma_i$ matrices which form the Clifford algebra

$$\{\,\gamma_{}^i,\gamma_{}^j\,\}~=~2\delta^{ij}_{}\ ,~~~~~~i,j~=1,\dots,9\ .$$
Supersymmetry is generated by the sixteen real supercharges 
$$
{\cal Q}^{a}_{\pm}={\cal Q}_{\pm}^{a\,*}\ ,
\nonumber$$
 which satisfy 
\bea
\nonumber\{{\cal Q}^a_+,{\cal Q}^b_+\}\,=\,\sqrt{2}\,p^+\delta^{ab}\, \\
\{{\cal Q}^a_-,{\cal Q}^b_-\}\,=\,\frac{{\vec p}\cdot{\vec p}}{\sqrt{2}\,p^+}\delta^{ab}\ ,\\
\nonumber\{{\cal Q}^a_+,{\cal Q}^b_-\}\,=\,-(\gamma_i)^{ab}p^i ,
\eea
and  transform  as Lorentz spinors
\bea
~[M^{ij},{\cal Q}^a_\pm]&=&\frac i2(\gamma_{}^{ij}{\cal Q}_\pm)^a\ ,\qquad 
~[M^{+-},{\cal Q}^a_\pm]~=~\pm\frac i2{\cal Q}_\pm^a\ ,\\
~[M^{\pm i},{\cal Q}^a_\mp]&=&0\ ,\qquad 
~[M^{\pm i},{\cal Q}^a_\mp]~=~\pm\frac i{\sqrt{2}}(\gamma_{}^i{\cal Q}_\pm)^a\ .
\eea
A very simple representation of the $11$-dimensional super-Poincar\'e generators can be constructed,
in terms of sixteen anticommuting real $\chi$'s and their derivatives, which transform as the spinor of $SO(9)$, as
\bea
{\cal Q}_+^{a}~=~\partial_{\chi^a}+\frac{1}{\sqrt{2}}p^+\chi^a\ ,\qquad {\cal Q}_-^a~=~
-\frac{p_{}^i}{p^+}\;\left(\gamma_{}^i\,{\cal Q}_+\right)^a_{}\ ,\eea

\bea
\nonumber~M^{ij}&=&x^ip^j-x^jp^i-\frac{i}{2}\chi\;\gamma_{}^{ij}\partial_\chi\ ,\\
\nonumber~M^{+-}&=&-x^-p^+ -\frac{i}{2}\chi\;\partial_\chi \ ,\\
\nonumber~M^{+i}&=&-x^ip^+\ ,\\
~M^{-i}&=&x^-p^i-\frac{1}{2}\{x^i,P^-\}+\frac {ip^j}{2p^+}\chi\gamma_{}^i\gamma_{}^j\partial_\chi  \ .
\eea
The light-cone little group transformations are  generated by   

$$
S^{ij}_{}~=~-\frac i2\chi\,\gamma^{ij}_{}\,\partial_\chi\ ,$$
which satisfy the  $SO(9)$ Lie algebra. To construct its spectrum,  we  write the supercharges in terms of eight complex Grassmann variables

$$
\theta^\alpha~\equiv~\frac{1}{\sqrt{2}}\left(\chi^\alpha+i\chi^{\alpha+8}\right)\ ,\qquad \overline\theta^\alpha ~\equiv~\frac{1}{\sqrt{2}}\left(\chi^\alpha-i\chi^{\alpha+8}\right)\ ,\label{choice}$$
and 

$$
\frac{\partial}{\partial\theta^\alpha}~\equiv~\frac{1}{\sqrt{2}}\left(\frac{\partial}{\partial\chi^\alpha}-i
\frac{\partial}{\partial\chi^{\alpha+8}}\right)\ ,\qquad 
\frac{\partial}{\partial\overline\theta^\alpha}~\equiv~
\frac{1}{\sqrt{2}}\left(\frac{\partial}{\partial\chi^\alpha}+i\frac{\partial}{\partial\chi^{\alpha+8}}\right)\ ,$$
where $\alpha=1,2,\dots,8$. The eight complex $\theta$  transform as the $({\bf 4}\, ,\, {\bf 2})$,   and $\overline\theta$ as the $(\overline{\bf 4}\,,\, {\bf 2}) $ of the $SU(4)\times SU(2)$ subgroup of $SO(9)$.  The eight complex supercharges 

\bea
{\bf Q}_+^\alpha
&\equiv &\frac{1}{\sqrt{2}}\left({\cal Q}_+^\alpha +i{\cal Q}_+^{\alpha+8}\right)~=~\frac{\partial}{\partial\overline\theta^\alpha}+
\frac{1}{\sqrt{2}}p^+\theta^\alpha\ ,\\
 {\bf Q}_+^{\alpha\,\dagger}
&\equiv&\frac{1}{\sqrt{2}}\left({\cal Q}_+^\alpha -i{\cal Q}_+^{\alpha+8}\right)~=~\frac{\partial}{\partial\theta^\alpha}+\frac{1}{\sqrt{2}}p^+\overline\theta^\alpha\ ,
\eea
 satisfy

\be
\{\,{\bf Q}_+^{\alpha}\,,\,{\bf Q}_+^{\beta\,\dagger}\,\}~=~\sqrt{2}\,p^+\,\delta^{\alpha\beta}\ .
\ee
They act irreducibly on  chiral superfields  which are annihilated by the covariant derivatives

$$
\left(\frac{\partial}{\partial\overline\theta^\alpha}-
\frac{1}{\sqrt{2}}p^+\theta^\alpha\right)\,\Phi(y^-,\theta)~=~0\ ,$$
where

$$
y^-~=~x^--\frac{i\theta\overline\theta}{\sqrt{2}}\ .$$
Expansion of the superfield in powers of the eight complex $\theta$'s yields $256$ components, with the following $SU(4)\times SU(2)$ properties
\bean
1~~~&\sim&~~~({\bf 1},{\bf 1})\ ,\\
\theta~~~&\sim&~~~({\bf 4},{\bf 2})\ ,\\
\theta\theta~~~&\sim&~~~({\bf 6},{\bf 3})\oplus({\bf 10},{\bf 1})\ ,\\
\theta\theta\theta~~~&\sim&~~~({\bf \overline {20}},{\bf 2})\oplus({\bf \overline {4} },{\bf  4})\ ,\\
\theta\theta\theta\theta~~~&\sim&~~~({\bf 15 },{\bf 3 })\oplus({\bf 1 },{\bf  5})\oplus({\bf 20' },{\bf 1 })\ ,\eean
and the higher powers yield the conjugate representations by duality. These make up the three $SO(9)$ representations of $N=1$ supergravity

\bean
{\bf 44}~&=&~({\bf 1  },{\bf 5 })\oplus({\bf 6 },{\bf  3})\oplus({\bf 20' },{\bf  1})\oplus({\bf 1 },{\bf  1})\ ,\\
{\bf 84}~&=&~({\bf 15  },{\bf 3 })\oplus({\bf \overline {10} },{\bf  1})\oplus({\bf 10 },{\bf  1})\oplus({\bf 6 },{\bf  3})\oplus({\bf 1 },{\bf  1})\ ,\\
{\bf 128}~&=&~({\bf 20  },{\bf 2 })\oplus({\bf \overline {20} },{\bf  2})\oplus({\bf 4 },{\bf  4})\oplus({\bf \overline {4} },{\bf  4})\oplus({\bf 4 },{\bf  2})\oplus({\bf \overline {4} },{\bf  2})\ .\eean
For future reference we note the $SU(4)\times SU(2)$ Dynkin weights of the $\theta$s, using the notation $(a_1,a_2,a_3;\,a)$,  

\bean
\theta^1&\sim(1,0,0;\, 1)\ ,&\theta^8\sim(1,0,0;\, -1)\ ,\\ 
\theta^4&\sim(-1,1,0;\, 1)\ ,&\theta^5\sim(-1,1,0;\, -1)\ ,\\
\theta^7&\sim(0,-1,1;\, 1)\ ,&\theta^2\sim(0,-1,1;\, -1)\ ,\\ 
\theta^6&\sim(0,0,-1;\, 1)\ ,&\theta^3\sim(0,0,-1;\, -1)\ , 
\eean 
which enables  us to find the  highest weights of the supergravity representations

\bean {\bf 44}~~&:&~~~\theta^1\theta^4\theta^5\theta^8~=~(0,2,0;\,0)~~\sim~~({\bf 20'}\,\,{\bf 1})\\
{\bf 84}~~&:&~~~\theta^1\theta^8~~~~~~~=~(2,0,0;\,0)~~\sim~~({\bf 10}\,,{\bf 1}\,)\\
{\bf 128}~~&:&~~~\theta^1\theta^4\theta^8~~~~=~(1,1,0;\,1)~~\sim~~({\bf 20}\,,{\bf 2}\,)\ ,\eean
together with their $SU(4)\times SU(2)$ properties. All other states are generated by acting on these highest weight states with the lowering operators. The highest weight chiral superfield that describes $N=1$ supergravity in eleven dimensions is simply

\be
\Phi~=~\theta^1\theta^8\,h(y^-,\vec x)~+~\theta^1\theta^4\theta^8\,\psi(y^-,\vec x)~+~\theta^1\theta^4\theta^5\theta^8\,A(y^-,\vec x)\ ,\ee
which summarizes the spectrum of the super-Poincar\'e algebra in eleven dimensions of either a free  field theory or a free superparticle. All other states are obtained by applying the $SO(9)$ lowering operators.
\vskip 1cm

Since the little group generators  act on a $256$-dimensional space, we can express them in terms  of sixteen $(256\times 256)$ matrices, $\Gamma^a_{}$, which   satisfy the Dirac algebra

\be\{\, \Gamma_{}^a\,,\, \Gamma_{}^b\,\}~=~\,2\delta_{}^{ab}\ .\ee 
 This leads to an  elegant representation of the $SO(9)$ generators

\be
S^{ij}=-\frac {i}{4}(\gamma^{ij})^{ab}\, \Gamma_{}^a\, \Gamma_{}^b~\equiv~-\frac{i}{2}f^{\,ij\,a\,b} \Gamma_{}^a\, \Gamma_{}^b\ . \label{S_ij} \ee
The coefficients

$$ f^{\,ij\,a\,b}~\equiv~ \frac{1}{2}(\gamma_{}^{ij})^{ab}\ ,$$
naturally appear in the commutator between the generators of $SO(9)$ and any spinor operator  $T_{}^a$, as 

\be[\,T_{}^{ij}\,,\,T_{}^a\,]~=~\frac{i}{2}\left(\gamma_{}^{ij}\,T\right)^a~=~if^{\,ij\,a\,b}\,T_{}^b\ .\ee
But there is more to it, the $(\gamma_{}^{ij})^{ab}$ can also be viewed as structure constants of a Lie algebra. Manifestly antisymmetric  under $a\leftrightarrow b$,  they can appear  in the commutator of two spinors into the $SO(9)$ generators   

\be [\,T_{}^a\,,\,T_{}^b\,]~=~ \frac {i}{2}(\gamma^{ij})^{ab} \,T_{}^{ij} ~=~\,f^{\,a\,b\,ij}\,T_{}^{ij}\ ,\ee
and one easily checks that  they satisfy the Jacobi identities. Remarkably, the $52$ operators  $T^{ij}$ and $T^a$ generate the exceptional Lie algebra $F_4$, showing explicitly how an exceptional Lie algebra appears in the light-cone formulation of supergravity in eleven dimensions! 

\subsection{Character Formula}
As we have seen, the supergravity degrees of freedom are labeled by the light-cone little group  $SO(9)$ acting on the  transverse plane indices, as $h_{(ij)}\sim (2000)$, $A_{[ijk]}\sim (0010)$, $\Psi_{i\,\alpha}\sim (1001)$, shown here in Dynkin's notation. 
 
Their group-theoretical properties are summarized in the following table 
\hskip 2cm
\begin{center}
\begin{tabular}{|c|c|c|c|}
\hline
$~{\rm irrep}~$& $(1001)$&$ (2000)$ & $(0010)$   \\
 \hline \hline         
$~D~ $&$  128$ & $ 44$ & $ 84$  \\
 \hline    
$~I_2~$& $256$& $88$ & $168$  \\
\hline
$~I_4~$& $640$& $232$ & $408$ \\                                                            
 \hline
$~I_6~$&$1792$& $712$ &$1080$\\ 
\hline
$~I_8~$&$5248$& $2440$ &$3000$\\ 
\hline
 \end{tabular}\end{center}
\vskip 0.3cm
where $D$ is the dimension of the representation, and $I^{}_n$ are the Dynkin indices of the representations, related to the four  Casimir operators of $SO(9)$. We note that the dimension and Dynkin indices of the fermion is the sum over those of the bosons, except for $I_8$, indicating that these three representations have much in common. The failure of the sum rule for $I_8$ can be traced to the lack of renormalizability of the theory~\cite{CURTRIGHT}. It is understood as the bosonic and fermionic representations stem from the two spinor representations of $SO(16)$ which differ only in their highest order invariant.

It is a remarkable fact that  the supergravity fields  are the  first of an infinite number of triplets~\cite{PR} of $SO(9)$ representations which display the same group-theoretical relations: equality of dimension and of all Dynkin indices except $I_8$ between one representation and the sum of the other two. Quantum theories of these  Euler triplets may  have very interesting divergence properties, as these numbers typically occur in higher loop calculations, and such equalities usually increase the degree of divergence, and the failure of the equality for $I_8$ is probably related to the lack of renormalizability of the theory~\cite{CURTRIGHT}. 

This mathematical fact has been traced to a  character formula~\cite{GKRS} related to the three equivalent embeddings of $SO(9)$ into $F_4$! This character formula is given by  

$$
V_{\lambda}\,\otimes\,S^+_{}\,-\,V_{\lambda}\,\otimes\,S^-_{}~=~\sum_{c }\,{\rm sgn}(c)\,U_{c\bullet\lambda}\ .$$
On the left-hand side,  $V_\lambda$ is a representation of $F_4$ written in terms of its $SO(9)$ subgroup, $S^\pm$ are the two spinor representations of $SO(16)$ written in terms of its anomalously embedded subgroup $SO(9)$, $\otimes$ denotes the normal Kronecker product of representations, and the $-$ denotes the naive substraction of representations. On the right-hand side, the sum is over  $c$, the  elements of the Weyl group  which map the Weyl chamber of $F_4$ into that of $SO(9)$.  In this case there are three elements, the ratio of the orders of the Weyl groups (it is also the Euler number of the coset manifold), and $U_{c\bullet\lambda}$ denotes the $SO(9)$ representation with highest weight $c\bullet\lambda$,  where

$$c\bullet\lambda~=~c\,(\lambda+\rho^{{}^{}}_{F_4})-\rho^{{}^{}}_{SO(9)}\ ,$$ 
and the $\rho$'s are the sum of the fundamental weights for each group, and ${\rm sgn}(c)$ is the index of $c$. Thus to each $F_4$ representation corresponds a triplet, called Euler triplet. The supergravity case is rather trivial as 

$$SO(16)\,\supset\,SO(9)\ ,\qquad S^+_{}\,\sim\,{\bf 128}={\bf 128}\ ,\qquad S^-_{}\,\sim\,{\bf 128}'={\bf 44}\,+\, {\bf 84}\ ,$$
and the character formula reduces to the  truism

$${\bf 128}\,-\,{\bf 44}\,-\, {\bf 84}~=~{\bf 128}\,-\,{\bf 44}\,-\, {\bf 84}\ .$$
This construction yields the general form of the Euler triplets: the Euler triplet corresponding to the $F_4$ representation $[\,a_1\,a_2\,a_3\,a_4\,]$ is made up of the following three $SO(9)$ representations listed in order of increasing dimensions:

$$
 (2+a_2+a_3+a_4,a_1,a_2,a_3)\ ,~ (a_2,a_1,1+a_2+a_3,a_4)\ ,~(1+a_2+a_3,a_1,a_2,1+a_3+a_4)$$
The spinor representations appear with odd entries in the fourth place. Euler triplets with the largest spinor and two bosons must have both $a_3$ and $a_4$ even or zero. 

Since the Dynkin indices of the product of two representations satisfy the composition law

$$I_{}^{(n)}[\lambda\otimes \mu]~=~d^{}_\lambda\,I_{}^{(n)}[\mu]+d_\mu^{}\,I_{}^{(n)}[\lambda]\ ,$$
it follows that the deficit in $I^{(8)}$ is always proportional to  

$$d_\lambda(I^{(8)}_{S^+}-I^{(8)}_{S^-})$$
where $d_\lambda$ is the dimension of the $F_4$ representation that generates it. This yields the hope that by summing over an infinite set of representations of $F_4$, it might be possible to erase the deficit. Such a theory would presumably be finite.

\subsection{The Kostant Operator}
This character formula can be viewed as the  index formula of a  Dirac-like operator~\cite{KOS} formed over the  coset $F^{}_4/SO(9)$. This coset is the sixteen-dimensional Cayley projective plane, over which we introduce the previously considered Clifford algebra    

\be
\{\, \Gamma_{}^a\,,\, \Gamma_{}^b\,\}~=~2\,\delta_{}^{ab}\ ,~~a,b=1,2,\dots, 16\ ,\ee
generated by $(256\times 256)$ matrices. The Kostant equation is defined as 

\be
\thru {\cal K}\,\Psi~=~\sum_{a=1}^{16}\, \Gamma_{}^a\,T^{a}_{}\,\Psi~=~0\ ,\ee
where $T_a$ are $F_4$ generators not in $SO(9)$, with  commutation relations

\be
[\,T_{}^a\,,\,T_{}^b\,]~=~i\,f^{\,ab\,ij}_{}\,T^{ij}_{}\ .\ee
Although it is taken over a compact manifold, it has non-trivial solutions. To see this, we rewrite its square as the difference of positive definite quantities, 

\be
\thru {\cal K}\, \thru {\cal K}\,~=~C^{2}_{F_4}-C^{2}_{SO(9)}+72\ ,\ee
where 

$$
C^{2}_{F_4}~=~\frac 12\,T_{}^{ij}\,T_{}^{ij}+T_{}^a\,T_{}^a\ ,$$
is the $F_4$ quadratic Casimir operator, and

$$
C^{2}_{SO(9)}~=~\frac 12\,\left(T_{}^{ij}-if_{}^{ab\,ij}\,\widetilde\Gamma_{}^{ab}\right)^2\ ,$$
 is the quadratic Casimir for the sum

$$
L_{}^{ij}~\equiv~T_{}^{ij}+S_{}^{ij}\ ,$$
where $S^{ij}$ is the previously defined $SO(9)$ generator which acts on the supergravity fileds.  We have also used  the quadratic Casimir on the spinor representation

$$
 \frac 12\,S_{}^{ij}\,S_{}^{ij}~=~72\ .$$
Kostant's operator commutes with the sum of the generators,

$$
[\,\thru {\cal K}\,,\,L_{}^{ij}\,]~=~0\ ,$$
allowing  its solutions to be labelled by $SO(9)$ quantum numbers. 

The same construction of Kostant's operator applies to all  equal rank embeddings, and its trivial solutions display supersymmetry~\cite{GKRS,BR,BR2,PR2}. In particular we note the cases $E_6/SO(10)\times SO(2)$, with Euler number $27$, $E_7/SO(12)\times SO(3)$ with Euler number $63$, and $E_8/SO(16)$, where the Euler triplets contain $135$ representations~\cite{PR}. These cosets with dimensions $32\ ,64$, and $128$ could be viewed as complex, quaternionic and octonionic Cayley plane~\cite{ATIYAH}.

\subsection{Solutions of Kostant's Equation}
For every representation of $F_4$,  $[a_1,a_2,a_3,a_4]$, there is one Euler triplet solution of Kostant's equation containing the three $SO(9)$ representations 

$$
 (2+a_2+a_3+a_4,a_1,a_2,a_3)\ ,~~(a_2,a_1,1+a_2+a_3,a_4)\ ,~~(1+a_2+a_3,a_1,a_2,1+a_3+a_4)\ .
$$
The trivial solution with   $a_1=a_2=a_3=a_4=0$,  yields the  $N=1$ supergravity multiplet in eleven dimensions, $(2000)\oplus(0010)\oplus(1001)$. In our representation, the highest weight solution are $\theta^1\theta^4\theta^5\theta^8\ , \theta^1\theta^8$, and $\theta^1\theta^4\theta^8$, described by the chiral superfield 

\be
\Phi_{0000}^{}~=~\theta^1\theta^8\,h(y^-,\vec x)~+~\theta^1\theta^4\theta^8\,\psi(y^-,\vec x)~+~\theta^1\theta^4\theta^5\theta^8\,A(y^-,\vec x)\ .\ee
The rest of the solutions contain ``internal" variables on which $F_4$ acts, in exact analogy with the $z_i$ of the previous section. A particularly convenient Schwinger-like representation $F_4$ can be found in Appendix C. It introduces three sets of ``internal" bosonic coordinates,  $u_i$ which transform as an $SO(9)$ vector, and  $\zeta_a$ which transform as an $SO(9)$ spinor. Curiously, an $SO(9)$-scalar coordinate $u_0$ required in the construction of the generators does not appear in the solutions of Kostant's equation. These coordinates span the  fundamental $\bf 26$ representation of $F_4$.  Note the appearance of a bosonic coordinates with spinorial properties, in apparent violation of the spin-statistics connection. 

In general, the  highest weight solutions appear in the form of $f(u_i,\zeta_a)\,\Theta(\theta)$,   where both $f(u_i,\zeta_a)$ and $\Theta(\theta)$ are the highest weights  $SO(9)$ states  with respect to $T_{ij}$ and $S_{ij}$. The solutions have the quantum numbers of the algebra generated by their sum  $L_{ij}=S_{ij}+T_{ij}$, which commutes with Kostant's operator. $\Theta(\theta)$ is one of the three polynomials, $\theta^1\theta^4\theta^5\theta^8\ , \theta^1\theta^8$, or $\theta^1\theta^4\theta^8$. 

It suffices to find the highest weight solutions, as all other solutions are obtained by the action of the four lowering $SO(9)$ operators. The highest weight solutions corresponding to each fundamental representation of $F_4$ are

\begin{enumerate} 

\item{$a_1=a_2=a_3=0,\, a_4\ge 1$} 

\noindent These representations require only one copy of the internal coordinates. The  highest weight solutions with $a_4=1$ are  given by

\be
\theta^1\theta^4\theta^5\theta^8\,w_4^{}\ , \qquad \theta^1\theta^8\,v_4^{}\ ,\qquad
\theta^1\theta^4\theta^8\,v_4^{}\ ,\ee
where 

\be w^{}_4~=~(u^{}_1+iu^{}_2)\ ,\qquad v_4^{}~=~(\zeta^{}_1+i\zeta^{}_9)\ ,\ee
 are the highest weights of $SO(9)$ representations $(1000)$ and $(0001)$, respectively.

\item{$a_1=a_2=a_4=0,\, a_3\ge 1$}

\noindent This case demands two copies. For $a_3=1$, the highest weight 
solutions are

\be
\theta^1\theta^4\theta^5\theta^8\,w^{}_3\ ,\qquad
\theta^1\theta^8\,v_3^{}\ ,\qquad
\theta^1\theta^4\theta^8\,w^{}_3\ ,\ee
where  
 
\be
w^{}_3~=~[\,u_1+iu_2\,,\,\zeta_1+i\zeta_9\,]\ ,\qquad  v_3^{}~=~[\,\zeta_1+i\zeta_9\,,\,\zeta_8+i\zeta_{16}\,]\ ,\ee are the highest weights of the $SO(9)$ 
representations $(1001)$ and $(0010)$ respectively, and

$$ [\,a\,,\,b\,]~\equiv~a^{[1]}b^{[2]}-a^{[2]}b^{[1]}\ ,$$
stands for  the determinant  of $2$ copies of $a$ and $b$ states.

\item{$a_1=a_3=a_4=0,\, a_2\ge 1$}

\noindent Here  three copies are needed. The $a_2=1$ highest weight solutions  are 

\be\theta^1\theta^4\theta^5\theta^8\,w^{}_2\ ,\qquad
\theta^1\theta^8w^{}_2\ ,\qquad
\theta^1\theta^4\theta^8w^{}_2\ ,\ee
where 

\be
w^{}_2~=~[\,u_1+iu_2\,,\zeta_1+i\zeta_9\,,\,\zeta_8+i\zeta_{16}\,]\ ,\ee
 is the highest weight of the $SO(9)$ representation $(1010)$, and $[\,a\,,\,b\,,\,c\,]$  is the determinant (antisymmetric product) of $3$ copies of $a,\,b$ and $c$ states. The same combination of internal variables appears for all three highest weight components.

\item{$a_2=a_3=a_4=0,\, a_1=1$}

\noindent The $F_4$ states are represented by antisymmetric products of two copies of $26$ states. The highest weight solutions are

\be
\theta^1\theta^4\theta^5\theta^8\,w^{}_1\ ,\qquad
\theta^1\theta^8\,w^{}_1\ ,\qquad
\theta^1\theta^4\theta^8\,w^{}_1\ ,
\ee
where 

\be w^{}_1~=~ [\,u_1+iu_2\,\,,u_3+iu_4\,]+[\,\zeta_1+i\zeta_9\,,\,\zeta_6-i\zeta_{14}\,]+[\,\zeta_8+i\zeta_{16}\,,\,\zeta_3-i\zeta_{11}\,]\ ,\ee
 is the highest weight of the $SO(9)$ representation $(0100)$. This is the algebraically most complicated case, but all three members of the triplet appear multiplied by the same combination of internal variables.
\end{enumerate}
\vskip .2cm

This last case implies that only three copies of $26$ oscillators suffice to generate all $F_4$  representations. It is not possible to construct the $[\,1\,0\,0\,0\,]$ state out of four copies of states in the $\bf 26$, and all representations of $F_4$ can be obtained by three copies of harmonic oscillator variables. 

Since the $T_{ij}$  do not alter the degree of homogeneity of  polynomials in $u_i$ and $\zeta_a$, all solutions are given by the functions $f(u_i,\zeta_a)$ as homogeneous polynomials of their variables. The general highest weight solutions  are  then

\bea
&&\theta^1\theta^4\theta^5\theta^8\,w_1^{a_1}\,w_2^{a_2}\,w_3^{a_3}\,w_4^{a_4}\ ,\nonumber\\
&&\theta^1\theta^8\,w_1^{a_1}\,w_2^{a_2}\,v_3^{ a_3}\,v_4^{ a_4}\ ,\nonumber\\
&&\theta^1\theta^4\theta^8\,w_1^{a_1}\,w_2^{a_2}\,w_3^{a_3}\,v_4^{a_4}\ .
\nonumber
\eea
All other states are generated by application of the four $SO(9)$ lowering operators. 
These explicit forms of the  solutions to Kostant's equation, as  products of a $\theta$ part and an internal part that depends polynomially on new variables, may suggest a physical interpretation.   Indeed, as the $\theta$ parts describe a superparticle in eleven dimensions, it is tempting to interpret states in the other Euler triplets as superparticles dressed with fields described by these new variables, vector coordinates $u^{[\kappa]}_i$ and twistor-like coordinates $\zeta^{[\kappa]}_a$.

\subsection{Super Euler Triplets}
 Any solution with odd powers $\zeta_a$ describes fermions ($SO(9)$ spinors) with Bose properties, in contradiction with the  spin-statistics connection. To avoid this conflict, we must restrict our attention to Euler triplets with even powers of $\zeta_a$; from the explicit forms of the solutions, this means Euler  triplets for which $a_3$ and $a_4$ are even, with no restrictions on $a_1$ and $a_2$.

The supergravity  Euler triplet  contains  equal number of fermions and bosons. Although none of the other  Euler triplets display space-time supersymmetry, we may ask which subclass  contains equal number of fermions ($SO(9)$ spinors) and bosons ($SO(9)$ tensors). Curiously, these are the same triplets as those required by spin-statistics, those for which $a_3$ and $a_4$ are even. 

In this section we concentrate on these special triplets which we call  Super Euler triplets (SETs) that share this one feature of supersymmetry, although they are not space-time supersymmetric by themselves. There are four different SET families: 

\begin{itemize}

\item The  simplest has $a_4$   even, and $a_3=a_2=a_1=0$. The general highest weight  superfield  in this class is given by (with generic spacetime fields)

\be
\Phi^{n}_{4}~=~ \phi(y^-,\vec x)\,\theta^1\theta^4\theta^5\theta^8\,w_4^{2n}+ A(y^-,\vec x)\,\theta^1\theta^8\,v_4^{2n}+\psi(y^-,\vec x)\,\theta^1\theta^4\theta^8\,v_4^{2n}\ .\ee
  This family requires  one new set of vector and twistor coordinates. For $n=1$ the internal coordinates generate  a symmetric second rank tensor represented by $(2000)$ coupled to gravity, and a four-form  $(0002)$, coupled to the three-form and Rarita-Schwinger fields of supergravity.

\item The case  $a_3\ne 0$ even, and $a_1=a_2=a_4=0$ requires two sets of extra coordinates, as its highest weight superfield is  
  
\be
\Phi^{n}_{3}~=~ \phi(y^-,\vec x)\,\theta^1\theta^4\theta^5\theta^8\,w_3^{2n}+ A(y^-,\vec x)\,\theta^1\theta^8\,v_3^{2n}+\psi(y^-,\vec x)\,\theta^1\theta^4\theta^8\,w_3^{2n}\ .\ee
The two sets of internal coordinates arrange themselves in the symmetric product of two three forms that couple to the supergravity three-form, and two RS spinors that couples to gravity and the original RS fields of supergravity.  
  
\item When only $a_2\ne 0$, all three supergravity multiplets are dressed by the same $(1010)$ representation, described by triple products of vector and two spinors. The highest weight superfield is now

\be
\Phi^{n}_{2}~=~ \left[\phi(y^-,\vec x)\,\theta^1\theta^4\theta^5\theta^8\,+ A(y^-,\vec x)\,\theta^1\theta^8\, +\psi(y^-,\vec x)\,\theta^1\theta^4\theta^8\right]\,w_2^n\ ,\ee
which requires three sets of coordinates.

\item Finally, if $a_1\ne 0$ only, the three supergravity states are dressed the same way by a combination of two vectors and two spinors with the quantum number of a 2-form, $(0100)$.
Although it is the most complicated in terms of the underlying coordinates, it is the simplest in terms of representations. 

\be
\Phi^{n}_{1}~=~\left[\phi(y^-,\vec x)\,\theta^1\theta^4\theta^5\theta^8\, +
 A(y^-,\vec x)\,\theta^1\theta^8\, +\psi(y^-,\vec x)\,\theta^1\theta^4\theta^8\right]\,
w^{n}_4\ .\ee
These SETs require  only two copies. This doubling may indicate the presence of $E_6$,  the complex extension of $F_4$. 

\end{itemize}
We note in passing that none of the solutions depend on  the singlet variable $u^{}_0$, which can be traced to the equation  

$$ \Gamma_{}^a\,\frac{\partial}{\partial \zeta_a}~=~0\ .$$

We  already know that space-time supersymmetry cannot be implemented on any one of these triplets, although they  may still be relativistic. For these Euler triplets to describe relativistic states, one must be able to implement their Poincar\'e transformations. In the light-cone form, the kinematic generators are already known, since the light-cone little group is generated by the sum $L^{ij}=T^{ij}+S^{ij}$, where $S^{ij}$ acts on the Grassmann variables and $T^{ij}$  on the internal variables. In order to determine the dynamic generators, one needs to find the mass operator. One of two possibilities arise

\begin{itemize}

\item The Euler triplet states are massless, and  there is no further addition to the light-cone Hamiltonian $P_{}^-$. These states represent massless higher spin particles, with well-known difficulties in implementing their interactions~\cite{DIFFICULT}. In this case an interacting theory would need to use an infinite number of Euler triplets to avoid these no-go theorems~\cite{VASILIEV}.   

\item If the excited Euler triplets  describe massive states,  a non-linear realization of  
the generators of the massive little group $SO(10)$ with the $SO(9)$ part given by $L^{ij}$ must exist.  This requires the construction of a transverse vector $L^i$, with the commutation relations
 
$$[\,L^i_{}\, ,\,L^j_{}\,]~=~iM^2_{}\,L^{ij}_{}\ ,$$
where $M^2$ is the mass squared operator which commutes with $L^{ij}$. If such an operator can be found, one easily builds the  light-cone boosts which satisfy the Poincar\'e algebra. For strings and superstrings, the $L^i$ are cubic in the oscillators, and this commutation relation  works only in the right number of dimensions. 
\end{itemize}

It is not possible to find such a non-linear representation of $SO(10)$ using only the degrees of freedom present in the Super Euler triplets, so that their relativistic description requires them to be massless.  To see this, consider the $SO(9)$ fermionic representation of the $\{a_1\,a_2\,a_3\,a_4\}$ Super Euler triplet, with Dynkin labels $(1+a_2+a_3,a_1,a_2,1+a_3+a_4)$ and  $a_3,a_4$ even. It can be realized in terms of fields with one spinor index and a tensor structure given by the partition 

$$\left[1+a_1+a_2+a_3+\frac{a_3+a_4}{2}\ ,\,a_1+a_2+\frac{a_3+a_4}{2}\ ,\,a_2+\frac{a_3+a_4}{2}\ ,\,\frac{a_3+a_4}{2}\right]\ ,$$
which has the first row of its Young tableau always larger than its second row. This representation is contained within an $SO(10)$ spinor-tensor which also has more in its first row than in its second. However such a representation, expressed in $SO(9)$ language,  also contains a partition in which  the excess in the first row is identified with the tenth direction, resulting in an  $SO(9)$ spinor-tensor with equal numbers in its first two rows. Since this representation is not in any super triplet, we must conclude that it is not possible to build massive relativistic Super Euler triplets without introducing new degrees of freedom.

At this stage, our approach still lacks an organizing principle to determine which set of  Euler triplets to include. A likely first restriction, from the  spin-statistics connection,  limits the multiplets to those with ($a_3,a_4$ even). A second one is to require that the deficit in $I^{(8)}$ be erased by summing over an infinite  set of Euler triplets. For instance an  infinite sum   of $F_4$ dimensions  involves a $24$th order polynomial over the Dynkin indices. Since it is an even power, it could vanish as $\zeta(-2m)=0$. We hope to come back to this point in a future publication. This also suggests the inclusion of $F_4$ inside a non-compact structure. 

We see that the Euler triplets generate the spectrum of a Poincar\'e covariant  object which has a ground state with supersymmetry!
This object would be described by its center of mass coordinates $x^-$ and $x_i$, and internal coordinates $u_i^{(\kappa)}\ ,\zeta_a^{(\kappa)}$, where $\kappa$ may run over three values at most. Contrast this with the superstring which is also described by its center of mass coordinates $x^-$, $x_m$, $m=1,\dots, 8$ and an infinite number of internal variables $x^{(n)}_m$, and anticommuting spinor variables $\zeta_\alpha^{(n)}$, $n=1,2,\dots \infty$, with $\alpha=1,\dots 8$. 

In this language, the Euler triplets emerge as much simpler than  superstrings, since they have a finite number of internal variables, although their internal spinor variables satisfy Bose commutations. 

Could these  label the end points of an open  string in the zero tension limit?

Could this new internal space be generated by  the degrees of freedom of the Exceptional Jordan Algebras?

\section{The Action for a Kostant Supermultiplet}
In the previous sections we have  described the free system of states which are subject to  Kostant's equation, using the light-cone formalism. The action that describes  these states as a set of point-like particles is  gauge-fixed. It would be very interesting to find the gauge-invariant action, but we have so far failed to do so. Even the gauge-fixed action is interesting as it requires Kostant's equation 
as a gauge condition. 

We write  this action in the light cone frame, in a generic form to cover the case we have analyzed. The superspace dynamical coordinates
will be $x^i$ with $i=1,2,...d-2$ and the spinorial ones $S^a$ with $a=1,...16$ for $d=11$ and correspondingly for other values of $d$. The internal ones will be $z^A$ and ${\overline z}_A$, where $A=1,2,3$ for the case of $SU(3)/SU(2)\times U(1)$ and $A=A_n$, where $A=1,...26$ and $n=1,..3$ (or $ 4$) for the case of $F_4/SO(9)$.

We then can  write the following  action

\be
S=\int d\tau (\frac{1}{2} ({\dot x}^i)^2 +i S^a{\dot S}^a + {{\dot z}^A}{{\dot {\overline z}}_A} + \lambda S^a T^a   ),
\ee
where $T^a$ are the generators for the quotient space (see the appendices), where we have exchanged $\frac{\partial}{\partial z}$ with $i \dot {\overline z}$ and the same for the complex conjugate

All coordinates are functions of $\tau$. The Lagrange multiplier $\lambda$ gives the Kostant constraint as an equation of motion. In Dirac's language the constraint arises since the momentum conjugate to $\lambda$, $\pi_{\lambda}$ is $0$. This is a constraint on the phase space and signals a gauge invariance. A typical gauge choice that we will make in the sequel at the level of the equations of motion is $\lambda=0$. It seems difficult to fix the gauge and put it back to the action as we have done with the local symmetries that allowed us to just use $x^i$ and $S^a$ as dynamical coordinates. We have fixed a local reparametrization symmetry to eliminate $x^-$ and $p^-$. We believe that some kind of $\kappa$-symmetry has allowed us to eliminate half a spinor. Presumably another bosonic symmetry has allowed us to eliminate $z$-variables down to the ones above in the action. In the gauge where $\lambda =0$ the equations of motion are

\be
{\ddot x}^i =0\ ,\qquad 
{\dot S}^a=0\ ,
\ee

\be
{\ddot z}^a =0\ ,\qquad
{\ddot {\overline z}}_A =0\ ,
\ee

\be
S^a T^a =0
\ee
The canonical commutation relations are

\be
[x^i,p^j] = i{\delta}^{ij}\ ,\qquad
\{S^a,S^b\}={\delta}^{ab}
\ee

\be
[z^A,{p_z}_B] = i{\delta}{^A}{_B}\ ,\qquad
[{\overline z}_A,{{\overline {p}_z}}^B] = i{\delta}{_A}^{B},
\ee
where the conjugate momenta are

\be
p^i= {\dot x}^i\ ,\qquad
{p_z}_A = {\dot {\overline z}}_A\ ,\qquad
{{\overline p}_z}^A = {\dot z}^A\ .
\ee
If we insert the $p_z$ for $\dot z$ in $T^a$ we get the form of the qoutient space generators.

We can represent the $S^a$'s in two different ways. In the first case we just choose 
\be 
S^a = \frac{1}{\sqrt 2}\gamma^a.
\ee

The second case we illustrate for the case when $a=1,2$. We choose 

\be
S^1+iS^2= \frac{\partial}{\partial\overline \theta}+\theta
\ee
and
\be
S^1-iS^2= \frac{\partial}{\partial \theta}+\overline \theta
\ee

We have now introduced too many degrees of freedom so we have to impose the chiral constraint. We have put $p^+=1$ in all these arguments to streamline it. We can then represent our particle on wave functions which are functions of a commuting set of the phase space coordinates, ie. $x^+$, $p^+$, $x^i$, $\theta$, $\overline \theta$, $z^A$ and ${\overline z}_A$, which are chiral and are subject to the Kostant constraint.

We notice that the two constraints, the chiral one and the Kostant one come from very different causes. The chiral condition comes from using a too big space while the Kostant constraint comes from a gauge invariance.

The action is invariant under the full Poincare algebra where the $SO(9)$
 part includes the $z$-dependence. The kinetic term for the $z$'s has the full $F_4$ invariance, but this symmetry is broken by the Kostant term.

Suppose we now try to introduce back the full gauge invariance. We can introduce a reparametrization invariance by using an einbein, but if we introduce a $\kappa$-invariance this will affect the Kostant term too, since we then have to double the spinor. What happens then to the Kostant term? Is there a group with a quotient space which is $32$-component. We know of the complex Cayley plane $E_6/SO(10)\times SO(2)$, but we have so far not been able to complete this argument.

 \vfill\eject
\setcounter{equation}{0}
\noindent{\LARGE\bf  Appendix A:}
\vskip .5cm
\noindent{\LARGE\bf  $SU(3)$ Oscillator Representations }
\vskip 1cm
\noindent
Schwinger's celebrated representation of $SU(2)$  generators  in terms of one doublet of harmonic oscillators has been  extended to other Lie algebras~\cite{FULTON}. The generalization involves several sets of harmonic oscillators, each spanning  the fundamental representations. Thus  $SU(3)$ is generated by two sets of  triplet harmonic oscillators, one transforming as a triplet the other as an antitriplet. Its generators are given by

$$ T^{}_1+iT^{}_2=z^{}_1\part2-\overline z^{}_2\partb1\ ,\qquad 
T^{}_1-iT^{}_2=z^{}_2\part1-\overline z^{}_1\partb2\ ,$$
$$T^{}_4+iT^{}_5=z^{}_1\part3-\overline z^{}_3\partb1\ ,\qquad 
T^{}_4-iT^{}_5=z^{}_3\part1-\overline z^{}_1\partb3\ ,$$
$$ T^{}_6+iT^{}_7=z^{}_2\part3-\overline z^{}_3\partb2\ ,\qquad 
T^{}_6-iT^{}_7=z^{}_3\part2-\overline z^{}_2\partb3\ ,$$
and

$$ T^{}_3=\frac{1}{2}(
z^{}_1\part1-z^{}_2\part2-\overline z^{}_1\partb1+\overline z^{}_2\partb2)\ ,$$
$$ T^{}_8=\frac{1}{2\sqrt{3}}(
z^{}_1\part1+z^{}_2\part2-\overline z^{}_1\partb1-\overline z^{}_2\partb2-2
z^{}_3\part3+2\overline z^{}_3\partb3)\ ,$$
where we have defined

$$\part1\equiv\pa1\ ,\qquad \partb1\equiv\pab1\ ,~{\rm etc.}\ .$$
These  act as hermitian operators on holomorphic functions of
$z_{1,2,3}^{}$ 
and $\overline z_{1,2,3}^{}$, normalized with respect to the inner product

$$ (f,g)\equiv~\int d^3zd^3\overline z~e^{-\sum_i\vert z_i\vert^2}~
f^*(z,\overline z)~g(z,\overline z)\ .$$
Acting on the highest-weight states, the $SU(3)$ quadratic Casimir operator is 

$$ C_2^{SU(3)}\equiv\sum_{a=1}^8T^{}_aT^{}_a\Big\vert_{\rm
highest~weight}=
T^{}_3(T^{}_3+1)+T^{}_8(T^{}_8+\sqrt{3})\ .$$
The second Casimir operator is cubic, and of no concern here. Rather
than labelling the representations in terms of their eigenvalues, it is
more convenient to introduce the 
positive integer Dynkin labels $a_1$ and $a_2$. We have

$$ T^{}_3\vert~ a_1,a_2>=~\frac{a_1}{2}\vert~ a_1,a_2>\ ,\qquad
T^{}_8\vert~ 
a_1,a_2>~=~\frac{1}{2\sqrt{3}}(a_1+2a_2)\vert~ a_1,a_2>\ ,$$
so that

$$ C_2^{SU(3)}~=~(a^{}_1+a^{}_2)+\frac{1}{3}(a_1^2+a_1^{}a_2^{}+a^2_2)\ .$$
The highest-weight states of each $SU(3)$ representation are holomorphic 
polynomials of the form

$$ z^{a_1}_1\overline z_3^{a_2}\ ,$$
where $a_1,a_2$ are its  Dynkin indices:  all
representations 
of $SU(3)$ are homogeneous holomorphic polynomials.

Finally we note that the Casimir operator of the $SU(2)$ subalgebra is given by

\be\vec T\cdot\vec T~=~\frac{1}{4}D_\perp(D_\perp+2)\ ,\ee
where

$$D^{}_\perp~=~z^{}_1\part1+z^{}_2\part2+\overline z^{}_1\partb1+\overline z^{}_2\partb2\ ,$$
so that the spin of the $SU(2)$ representation is simply

\be
J~=~\frac{1}{2}D^{}_\perp\ .\ee
\vfill\eject
\setcounter{equation}{0}
\noindent{\LARGE\bf  Appendix B:}
\vskip .5cm
\noindent{\LARGE\bf Euler Triplet Ladder Operations}
\vskip .5cm

\noindent In section 2 we treated the highest weight states and showed how to go from one state to another. In this Appendix we will treat the general case. Start with the general form of the Euler triplet superfields
\bea\nonumber
\label{Gen}
\Phi_{\{a_1,a_2\}}^{}(y^-,x^i;\,{\bf z},{\bf \overline z},\,\theta_m)~=&~& 
\\\nonumber\\\nonumber
\psi^{0}_{\{a_1,a_2\}}(y^-,x^i;\,{\bf z},{\bf \overline z}) &+& \theta^{}_m \psi_{\{a_1,a_2\}}^m(y^-,x^i;\,{\bf z},{\bf \overline z})~ +~ 
\theta^{}_1 \theta^{}_2\psi_{\{a_1,a_2\}}^{12}(y^-,x^i;\,{\bf z},{\bf \overline z})\ ,\eea
where the components are homogeneous polynomials or order $a_1$ in the variables $\bf z$ and $a_2$ in the $\bf \overline z$. The simplest examples are 

\begin{itemize}
\item The ``trivial" $\{0,0\}$  Euler triplet, described by the  superfield

\be
\label{Phi1}
\Phi_{\{0,0\}}^{}(y^-,x^i,\theta_m)= \psi^{0}_{\{0,0\}}(y^-,x^i) + \theta^{}_m \psi_{\{0,0\}}^m(y^-,x^i) + 
\theta^{}_1 \theta^{}_2\psi_{\{0,0\}}^{12}(y^-,x^i)\ ,\ee
where the $\psi$'s  depend  only on 
the center of mass variables $y^-$, and $x_i$, cf. (\ref{Ch}), not on the internal $z$ variables. The kinematic $N=2$ supersymmetry 
acts on its components by means of the operators

\be
{\cal Q}_+^m\approx -2p^+\theta_m\ ,\qquad \overline{\cal
Q}_+^m\approx 
\frac{\partial}{\partial\theta_m}\ ,\ee
so that this superfield describes  the $N=2$ Hypermultiplet on  the light-cone.

\item The $\{1,0\}$ triplet components are linear polynomials in the $z_{i}$ 

\bea\nonumber
\psi_{\{1,0\}}^0&=&b^{}_0z_3^{}\ ,\qquad\qquad ~~\,\,  \psi_{\{1,0\}}^{12}~=~b^{}_{12}
z^{}_1+b^{'}_{12}z^{}_2\ ,\\\nonumber
 \psi_{\{1,0\}}^1&=&b^{}_1z^{}_1+b^{}_2z^{}_2\ ,\qquad
\psi_{\{1,0\}}^2~=~b^{}_2z^{}_1+b^{}_3z^{}_2\ ,\eea
where the $b$'s depend only on the center of mass variables.

\item Similarly, the $\{2,0\}$ Euler triplet is represented by the quadratic polynomials

\bea\nonumber
\psi_{\{2,0\}}^0&=&c^{}_0z_3^{2}\ ,\qquad\qquad\qquad\qquad ~~~~  \psi_{\{2,0\}}^{12}~=~c^{}_{12}
z^{2}_1+c^{'}_{12}z_1z^{}_2+c^{''}_{12}z^{2}_2\ ,\\\nonumber
 \psi_{\{2,0\}}^1&=&c^{}_1z^{2}_1+2c^{}_2z^{}_1z^{}_2+c^{}_3z^{2}_2\ ,\qquad
\psi_{\{2,0\}}^2~=~c^{}_2z^{2}_1+2c^{}_3z^{}_1z^{}_2+c^{}_4z^{2}_2\ ,\eea
with the $c$'s depending  on the center of mass variables as above. Note that while $\psi^0$ and $\psi^{12}$ are arbitrary polynomials of a given degree of homogeneity, some of the same coefficients appear in $\psi^1$ and $\psi^2$.
\end{itemize}

If we want to generate {\em all} states of an Euler triplet, then we need to consider complicated projection schemes that single out the requisite states. As an example, let us see how the $\{1,0\}$ and $\{2,0\}$ triplets are generated from the supersymmetric $\{0,0\}$. When applied to the lowest Euler triplet,  the 
$SU(2)$-doublet operator  
$z_i\,(1-P^{}_0)$, where $(1-P_0)$ is the projection operator onto the $\theta$-dependent terms, generates more than  the
states in $\Psi_{\{1,0\}}$. In order to single out the  requisite
combinations of $z$'s and $\theta$'s, we need  the  operator

\be
{\cal P}~=~\frac{1}{2}\Big(1+\sum_{i,j=1,2}~z^{}_i\theta^{}_j
\frac{\partial}{\partial z^{}_j}\frac{\partial}{\partial\theta^{}_i}\Big)\ .\ee
It carries no helicity and is an $SU(2)$ singlet, as we  can  see by rewriting it in the form
\be
{\cal P}~=~\frac{1}{2}\Big(1+2\vec S\cdot \vec T + \frac{1}{2}\theta_i\frac{\partial}{\partial\theta_i}(z_j\frac{\partial}{\partial z_j}+
\overline z_j\frac{\partial}{\partial \overline z_j})
\Big)\ .\ee
Acting on the terms linear in $\theta$'s, it reduces  to  

\be
{\cal P}~=~\frac{1}{2}+\vec S\cdot \vec T+  J/2\ ,\ee
while it is equal to one-half acting on the $1\ ,\, \theta_1\theta_2$ terms. With the help of 

\be
(\vec S\cdot\vec T)^2~=~-\frac{1}{2}\,\vec S\cdot\vec T+\frac{1}{3}\,\vec S\cdot\vec S\,J(J+1)\ ,\ee
we see that on the terms linear in $\theta$'s, 

\be
{\cal P}^2_{}~=~(J+\frac{1}{2})\,{\cal P}\ .\ee
so that
\be
{\cal M}~=~\frac{{\cal P}}{J+\frac{1}{2}}\ ,\ee
is a true projection operator on the $\theta$-linear terms. It is easy to check that 

\bea\nonumber
{\cal P}~z_1\theta^{}_1~=~z^{}_1\theta^{}_1\ ~~&;&~~ {\cal
P}~z^{}_2\theta^{}_2
~=~z^{}_2\theta^{}_2\ ;\\ {\cal
P}~(z^{}_2\theta^{}_1+z^{}_1\theta^{}_2)
&=&(z^{}_2\theta^{}_1+z^{}_1\theta^{}_2)\ ,\eea
while  the unwanted combination 

\be
{\cal P}~(z_2\theta_1-z_1\theta_2)~=~0\ ,\ee 
is annihilated. Multiplying the $\theta$-dependent superfield components in the
$SU(2)$ spin $j$ representation by  $z_i$,
produces two $SU(2)$ representations,  $j+1/2$ and 
$j-1/2$. The action of ${\cal P}$ is to project out the latter (like the well-known projection that appears in $L-S$ coupling problems). Hence
the $SU(2)$-doublet ``creation" operators   have the simple form

\be {\cal A}^{\dagger}_i~\equiv~ {\cal P}(1-P^{}_0) ~z^{}_i \ .\ee
Explicit computations yield 

\bea
{\cal A}^{\dagger}_0~\Psi_{\{0,0\}}~&=&~\psi^0_{\{0,0\}}z^{}_3\nonumber\\
{\cal A}^{\dagger}_1~\Psi_{\{0,0\}}~&=&~\psi^{1}_{\{0,0\}}z^{}_1\theta^{}_1+\frac{1}{2}\psi^{2}_{\{0,0\}}
(z^{}_1\theta^{}_2+z^{}_2\theta^{}_1)
+\psi^{12}_{\{0,0\}}z^{}_1\theta^{}_1\theta^{}_2
\ ,\nonumber\\
{\cal A}^{\dagger}_2~\Psi_{\{0,0\}}~&=&~\frac{1}{2}\psi^{1}_{\{0,0\}}(z^{}_1\theta^{}_2+z^{}_2
\theta^{}_1)+\psi^{2}_{\{0,0\}}z^{}_2\theta^{}_2
+\psi^{12}_{\{0,0\}}z^{}_2\theta^{}_1\theta^{}_2
\ ,\nonumber\eea
so that their action on  the lowest Euler triplet generates all the
states in $\{1,0\}$, although one of the states is generated in two ways. The same redundancy is seen in the  double application of the step-up operators  on the lowest Euler triplet,

\bea
({\cal A}^\dagger_0)^2~\Psi_{\{0,0\}}&=&\psi^{0}_{\{0,0\}}z^2_3\ ,\nonumber\\
({\cal A}^\dagger_1)^2~\Psi_{\{0,0\}}&=&\frac{3}{2}\psi^{1}_{\{0,0\}}z^{2}_1\theta^{}_1+\psi^{2}_{\{0,0\}}(z^2_1\theta^{}_2
+2z^{}_1z^{}_2\theta^{}_1)+\frac{3}{2}\psi^{12}_{\{0,0\}}z^2_1\theta^{}_1\theta^{}_2\ ,\nonumber\\~
({\cal A}^\dagger_2)^2~\Psi_{\{0,0\}}&=&\psi^{1}_{\{0,0\}}(z^2_2\theta_1^{}+2z^{}_1z^{}_2\theta^{}_2)
+\frac{3}{2}\psi^{2}_{\{0,0\}}z^2_2\theta^{}_2+\frac{3}{2}\psi^{12}_{\{0,0\}}z^2_2\theta^{}_1\theta^{}_2\ ,\nonumber\\
{\cal A}^\dagger_1~{\cal A}^\dagger_2~\Psi_{\{0,0\}}&=&
\frac{1}{2}\psi^{2}_{\{0,0\}}(z^2_2\theta_1^{}+2z^{}_1z^{}_2\theta^{}_2)+\psi^{1}_{\{0,0\}}(z_1^2
\theta^{}_2+2z^{}_1z^{}_2\theta^{}_1)+\frac{3}{2}\psi^{12}_{\{0,0\}}z^{}_1z^{}_2\theta^{}_1\theta_2^{}\
,\nonumber\\
{\cal A}^\dagger_2~{\cal A}^\dagger_1~\Psi_{\{0,0\}}&=&\frac{1}{2}\psi^{1}_{\{0,0\}}(z_1^2
\theta^{}_2+2z^{}_1z^{}_2\theta^{}_1)+\psi^{2}_{\{0,0\}}(z^2_2\theta_1^{}+2
z^{}_1z^{}_2\theta^{}_2)+\frac{3}{2}\psi^{12}_{\{0,0\}}z^{}_1z^{}_2\theta^{}_1\theta_2^{}\ .
\nonumber\eea
Only the states in the $\{2,0\}$ multiplet are generated, although the same states can be generated by different sets of operators. This construction  generalizes easily to all triplets of the form
$\{a_1,0\}$: acting on  any triplet $\Psi_{\{a_1,0\}}$, the step-up operators ${\cal A}_{0,1,2}^\dagger$  
yield all the states in  the three $SU(2)$ representations of the triplet $\{a_1+1,0\}$. A similar construction holds for the step-down operators.

It is also easy to construct the ladder operators that relate triplets of the form $\Psi^{}_{\{0,a_2\}}$. These are defined as 

\be {\overline {\cal A}}^{\dagger}_{12}~\equiv~  P_{12}\,{\bar z}^{}_3 \ ,\qquad {\overline {\cal A}}^{\dagger}_i~\equiv~ {\bar {\cal P}}\,(1-P^{}_{12})\,{\bar z}^{}_i \ ,\ee

where

\be {\bar {\cal P}}= (1- \epsilon_{ij}\epsilon_{kl}{\bar z}^{}_i\theta^{}_k\frac{\partial}{\partial {\bar z}^{}_l}
\frac{\partial}{\partial \theta^{}_j})\ .
\ee  
and the corresponding ``annihilation''operators 
\be 
{\overline {\cal A}}_{12}~\equiv~ P^{}_{12}\,{\frac{\partial}{\partial {\bar z}^{}_3}} \ ,\qquad
{\overline {\cal A}}_i~\equiv~ {\bar {\cal P}}(1-P^{}_{12})~{\frac{\partial}{\partial {\bar z}^{}_i}} \ .\ee

%$$
%\Psi^{}_{\{0,1\}}={\bar b}^{}_0{\bar z}_2 - {{\bar b}^{'}}_0{\bar z}_1+{\bar b}^{}_1{\bar z}^{}_2
%\theta^{}_1+{\bar b}^{}_2({\bar z}^{}_2
%\theta^{}_2-{\bar z}^{}_1\theta^{}_1)-{\bar b}^{}_3{\bar z}^{}_1\theta^{}_2+{\bar b}^{}_{12}{\bar z}{}_3\theta^{}_1\theta^{}_2 ,$$ 
%where the $\bar b$'s depend only on the center of mass variables as above.

We can define the operators that generate shadow supersymmetries in terms of these new ladder operators

\be
{\cal Q}^{~[i]~}_{a~~b}~\equiv~ {\cal A}^\dagger_a~\theta^{}_i~
{\cal A}^{}_b\ ,\ee
where $a,b=0,1,2$, and $i=1,2$, and their inverses
\be
\overline{\cal Q}^{~[i]~}_{a~~b}~\equiv~ {\cal A}^{\dagger}_a~\frac{\partial}{\partial\theta^{}_i~}
{\cal A}^{}_b\ .\ee
Similarly we define the operators

\be
{Q}^{~[i]~}_{a~~b}~\equiv~{\bar {\cal A}}^\dagger_a~\theta^{}_i~
{\bar {\cal A}}^{}_b\ ,\ee
where $a,b=0,1,2$, and $i=1,2$, and their inverses
\be
\overline{Q}^{~[i]~}_{a~~b}~\equiv~ {\bar {\cal A}}^{\dagger}_a~\frac{\partial}{\partial\theta^{}_i~}
{\bar {\cal A}}^{}_b\ .\ee
They operate in a similar fashion on $\Psi_{\{0,1\}}$ and indeed on the whole superfield. We can extend these shadow supersymmetries to operations that step up N steps 

\be
{\cal Q}^{~[i]~}_{a_1~a_2..a_N~~b_1~b_2...b_N}~\equiv~ {\cal A}^\dagger_{a_1}~{\cal A}^\dagger_{a_2}..{\cal A}^\dagger_{a_N}~\theta^{}_i~
{\cal A}^{}_{b_1}~{\cal A}^{}_{b_2}...{\cal A}^{}_{b_N}\ ,\ee
where $a_i,b_j=0,1,2$, and $i=1,2$, and their inverses
\be
\overline{\cal Q}^{~[i]~}_{a_1~a_2..a_N~~b_1~b_2...b_N}~\equiv~ {\cal A}^{\dagger}_{a_1}~ {\cal A}^{\dagger}_{a_2}... {\cal A}^{\dagger}_{a_N}~\frac{\partial}{\partial\theta^{}_i~}
{\cal A}^{}_{b_1}~{\cal A}^{}_{b_2}..{\cal A}^{}_{b_N}\ee
and similarly for the ones with $\bar {\cal A}$. The anticommutators between these operators will close to terms of the form $z_i\frac{\partial}{\partial { z}^{}_i}P_j$.

We  can also check the anticommutators $$\{{\cal Q}^{~[i]~}_{a_1~a_2..a_N~~b_1~b_2...b_N}, 
\overline{\cal Q}^{~[i]~}_{a_1~a_2..a_M~~b_1~b_2...b_M}\}$$ to see that they close to step operators. Al these operators generate an infinite superalgebra for which the superfield is a representation, although we have not been able to to write it in a compact form. It would be very important to find a finite subalgebra. However, there are some arguments against such an algebra since it would amount to the solve the same problem as the old one of finding supersymmetries with higher (half-integer) helicity operators than $1/2$.

\vfill\eject
\setcounter{equation}{0}
\noindent{\LARGE\bf  Appendix C:}
\vskip .5cm
\noindent{\LARGE\bf  $F_4$ Oscillator Representations }
\vskip 1cm
\noindent In this case, it turns out that all representations of the exceptional group $F_4$ are generated by three (not four~\cite{FULTON}) sets of oscillators transforming as  ${\bf 26}$. We give the construction for compact and one non-compact case.
\vskip .2cm
\noindent {\bf Compact Case}
\vskip .2cm
\noindent
We label each copy of $26$ oscillators
as $A^{[\kappa]}_0,\; A^{[\kappa]}_i,\; i=1,\cdots,9,\; B^{[\kappa]}_a,\; a=1,\cdots,16$, 
and their hermitian conjugates, and where $\kappa=1,2,3 $. 
Under $SO(9)$, the $A^{[\kappa]}_i$ transform as ${\bf 9}$, $B^{[\kappa]}_a$ transform as ${\bf 16}$, 
and $A^{[\kappa]}_0$ is a scalar. They satisfy the commutation relations of ordinary   harmonic oscillators

$$
[\,A^{[\kappa]}_i\,,\,A^{[\kappa']\,\dagger}_j\, ]~=~\delta^{}_{ij}\,\delta_{}^{[\kappa]\,[\kappa']}\ ,\qquad [\,A^{[\kappa]}_0\,,\,A^{[\kappa']\,\dagger}_0\, ]~=~\delta_{}^{[\kappa\,\kappa']}\ . 
$$
Note that the  $SO(9)$  spinor operators   satisfy Bose-like commutation relations

$$ \nonumber[\,B^{[\kappa]}_a\,,\,B^{[\kappa']\,\dagger}_b\, ]~=~\delta^{}_{ab}\,\delta_{}^{[\kappa]\,[\kappa']}\ .$$
The   generators $T_{ij}$ and $T_a$  

\begin{eqnarray}
T^{}_{ij}&=&-i\sum_{\kappa=1}^4\left\{\left(A^{[\kappa]\dag}_iA^{[\kappa]}_j-A^{[\kappa]\dag]}_jA^{[\kappa]}_i\right)+\frac 12\,B^{[\kappa]\dag}\,\gamma^{}_{ij} B^{[\kappa]}\label{t_ij}\right\}\nonumber\ ,\\
T_a&=&-\frac{i}{{2}}\sum_{\kappa=1}^4\left\{ (\gamma_i)^{ab}\left(A^{[\kappa]\dag}_iB^{[\kappa]}_b-B^{[\kappa]\dag}_bA^{[\kappa]}_i\right)-\sqrt{3}\left(B^{[\kappa]\dag}_aA^{[\kappa]}_0-A^{[\kappa]\dag}_0B^{[\kappa]}_a\right)\right\}\nonumber\ ,\label{t_a}
\end{eqnarray}
satisfy the $F_4$ algebra,

\bean
[\,T^{}_{ij}\,,T^{}_{kl}\,]&=&-i\,(\delta^{}_{jk}\,T^{}_{il}+\delta^{}_{il}\,T^{}_{jk}-\delta^{}_{ik}\,T^{}_{jl}-\delta^{}_{jl}\,T^{}_{ik})\ ,\\
~[\,T^{}_{ij}\,,T^{}_a\,]&=&\frac i2\,(\gamma^{}_{ij})^{}_{ab}\,T^{}_b\ ,\\
~[\,T^{}_a\,,T^{}_b\,]&=&\frac i2\,(\gamma^{}_{ij})^{}_{ab}\,T^{}_{ij}\ ,
\eean
so that the structure constants are given by

$$
f_{ij\,ab}~=~f_{ab\,ij}~=~\frac{1}{2}\,(\gamma^{}_{ij})_{ab}\ .$$
The last commutator  requires the Fierz-derived identity  

$$
\frac{1}{4}\,\theta\,\gamma^{ij}\,\theta\;\chi\,\gamma^{ij}\,\chi~=~3\,\theta\,\chi\;\chi\,\theta+\theta\,\gamma^i\,\chi\;\chi\,\gamma^i\theta\ ,$$
from which we deduce

$$
3\,\delta^{ac}\delta^{db}+(\gamma^i)^{ac}\,(\gamma^i)^{db}- (a\leftrightarrow b)~=~
\frac{1}{4}\,(\gamma^{ij})^{ab}\,(\gamma^{ij})^{cd}\ .$$
To satisfy these commutation relations, we have required  both $A_0$ and $B_a$ to obey Bose commutation relations (Curiously, if  both  are anticommuting, the $F_4$ algebra is still satisfied). One can just as easily use a coordinate representation of the oscillators by introducing real coordinates $u^{}_i$ which transform as transverse space vectors, $u^{}_0$ as scalars, and $\zeta^{}_a $ as space spinors which satisfy Bose commutation rules

\bean
A_i&=&\frac 1{\sqrt{2}}(u^{}_i+\partial_{u^{}_i})\ ,\qquad A^\dag_i=\frac 1{\sqrt{2}}(u^{}_i-\partial_{u^{}_i})\ ,\\
B_a&=&\frac 1{\sqrt{2}}(\zeta^{}_a+\partial_{\zeta^{}_a})\ ,\qquad B^\dag_a=\frac 1{\sqrt{2}}(\zeta^{}_a-\partial_{\zeta^{}_a})\ ,\\
A_0&=&\frac 1{\sqrt{2}}(u^{}_0+\partial_{u^{}_0})\ ,\qquad A^\dag_0=\frac 1{\sqrt{2}}(u^{}_0-\partial_{u^{}_0})\ .
\eean
Using square brackets $[\cdots]$ to represent the Dynkin label of $F_4$, and
round brackets $(\cdots)$ to represent those of $SO(9)$, we list some of the combinations which will be used for investigating the solutions of Kostant's equation

\bean
u^{}_1+iu^{}_2&\sim& [\,0~~\,0~~\,0~~\,1\,]~\sim~ (~\,1~~\,0~~\,0~~\,0~)~ \ , \\
u^{}_3+iu^{}_4&\sim& [\,1~~\,0~~\,0-\!\! 1\,]~\sim~ (-\!\! 1~~\,1~~\,0~~\,0~) \ , \\
\zeta^{}_1+i\zeta^{}_9&\sim &[\,0~~\,0~~\,1-\!\!1\,]~\sim~ (~\,0~~\,0~~\,0~~\,1~) \ , \\
\zeta^{}_8+i\zeta^{}_{16}&\sim& [\,0~~\,1-\!\!1~~\,0\,]~\sim ~(~\,0~~\,0~~\,1-\!\! 1~) \ , \\
\zeta^{}_3-i\zeta^{}_{11}&\sim& [\,1-\!\!1~~\,1~~\,0\,]~\sim~ (~\,0~~\,1-\!\! 1~~\,1~) \ , \\
\zeta^{}_6-i\zeta^{}_{14}&\sim& [\,1~~\,0-\!\!1~~\,1\,]~\sim~ (~\,0~~\,1~~\,0-\!\! 1~)  \ . 
\eean
Hence $u^{}_1+iu^{}_2$ and $\zeta^{}_1+i\zeta^{}_9$  are the highest weights of the $SO(9)$ representations $\bf 9$, and $\bf 16$, respectively.

\vskip .2cm
\noindent {\bf Non-Compact Case}
\vskip .2cm
\noindent With a slight modification we can use this representation for the non-compact form of the embedding of $F_4$ in $SO(25,1)$.

The non-compact algebra $SO(25,1)$ can be  generated by changing the sign of one of the oscillator commutation relations, requiring that 

\be
[\,{\cal A}^{}_M\,,\,{\cal A}^{\dag}_N\,]~=~\eta^{}_{MN}\ ,\ee
where 

$$\eta^{}_{00}~=~-1\ ,\qquad \eta^{}_{ij}~=~\delta^{}_{ij}\ ,\qquad \eta^{}_{ab}~=~\delta^{}_{ab}\ .$$
The non-compact $F_4$ subalgebra still has $SO(9)$ as a compact subalgebra realized by the same hermitian generators 

$$
\widehat T^{}_{ij}~=~-i\left(A^{\dag}_iA^{}_j-A^{\dag]}_jA^{}_i\right)-\frac i2\,B^{\dag}\,\gamma^{ij}_{}B^{}\ .$$
The spinor generators  are given by

$${\widehat T}^{}_a~=~-\frac{1}{\sqrt{2}}\left\{ (\gamma_i)^{ab}\left(A^{\dag}_iB^{}_b-B^{\dag}_bA^{}_i\right)-i\sqrt{3}\left(B^{\dag}_aA^{}_0-A^{\dag}_0B^{}_a\right)\right\}\ .$$
Note that they are no longer hermitian, still transform as $SO(9)$ spinors, but obey the ``wrong sign" commutator 

$$ [\,\widehat T^{}_a\,,\widehat T^{}_b\,]~=~-i(\gamma_{ij})^{}_{ab}\,\widehat T^{}_{ij}\ .$$
This corresponds to the particular non-compact form of $F_4$ with maximal $SO(9)$ compact subgroup.

Note that these are not simply the compact form multiplied by $i$. Because of the minus sign in the $A_0$ commutator, the $B_aA^\dagger_0$ part of $T_a$ gives a term like $B^\dagger\,B$ with the opposite sign from the compact case. This is good because we need to get a $\gamma^{ij}\gamma^{ij}$ term of the opposite sign to get the non-compact commutator, as these two are linked by 
the Fierz identity. This means that the $\gamma^i\gamma^i$ term must be the negative of the compact case; this is done by taking away the $i$ that multiplies the $A^iB^\dagger_b(\gamma)^i_{ab}$ terms in $T_a$, leading to the form above.

\vfill\eject
%\section {References}

\end{document}